\documentclass[fleqn, prd, nofootinbib, superscriptaddress]{revtex4-2}
\usepackage[utf8]{inputenc}
\usepackage[T1]{fontenc}
\usepackage[english]{babel}
\usepackage{amssymb, amsmath, graphicx, epsfig, bm, appendix, braket}
\usepackage[caption=false]{subfig}
\usepackage{xcolor}
\usepackage{hyperref}
\usepackage[capitalise]{cleveref}
\definecolor{blue_refs}{rgb}{0., 0., 0.85}
\hypersetup{ 
colorlinks = true,
linkcolor = blue_refs,
citecolor = blue_refs,		
filecolor = blue_refs,		
urlcolor = blue_refs
}		

\renewcommand{\d}{\mathrm{d}}
\newcommand{\e}{\mathrm{e}}

\newcommand{\sT}{{\scriptscriptstyle T}}
\newcommand{\sL}{{\scriptscriptstyle L}}

\newcommand{\pT}{\bm{p}_\sT}

\DeclareMathOperator{\tr}{Tr}

\def\nn{\nonumber}

\def\slash#1{\setbox0=\hbox{$#1$}               
        \dimen0=\wd0                            
        \setbox1=\hbox{/} \dimen1=\wd1          
        \ifdim\dimen0>\dimen1                   
        \rlap{\hbox to \dimen0{\hfil/\hfil}}    
        #1                                      
        \else              
        \rlap{\hbox to \dimen1{\hfil$#1$\hfil}} 
        /                                       
        \fi}                                    %

\begin{document}
\title{Spin asymmetries for \texorpdfstring{$C$}{\it C}-even quarkonium production as a probe of gluon distributions}

\author{Nanako Kato}
\email{nanako.kato@ca.infn.it}
\affiliation{Dipartimento di Fisica, Università di Cagliari, Cittadella Universitaria, I-09042 Monserrato (CA), Italy}
\affiliation{INFN, Sezione di Cagliari, Cittadella Universitaria, I-09042 Monserrato (CA), Italy}

\author{Luca Maxia}
\email{l.maxia@rug.nl}
\affiliation{Van Swinderen Institute for Particle Physics and Gravity, University of Groningen, Nijenborgh 4, 9747 AG Groningen, The Netherlands}

\author{Cristian Pisano}
\email{cristian.pisano@unica.it}
\affiliation{Dipartimento di Fisica, Università di Cagliari, Cittadella Universitaria, I-09042 Monserrato (CA), Italy}
\affiliation{INFN, Sezione di Cagliari, Cittadella Universitaria, I-09042 Monserrato (CA), Italy}

\begin{abstract}
Within the framework of transverse momentum dependent factorization in combination with nonrelativistic QCD, we study charmonium and bottomonium production in hadronic collisions. We focus on quarkonium states with even charge conjugation, for which the color-singlet production mechanism is expected to be also dominant in the small transverse momentum region, $q_\sT^2 \ll 4 M_{c,b}^2$. It is shown that the distributions of linearly polarized gluons inside unpolarized, longitudinally and transversely polarized protons contribute to the cross sections for scalar and pseudoscalar quarkonia in a very distinctive,  parity-dependent way, whereas their effects on higher angular momentum states are strongly suppressed. We derive analytical expressions for single and double spin asymmetries, which would allow for the direct extraction of the gluon transverse momentum dependent distributions, mirroring the phenomenological studies of the Drell-Yan processes aimed at the extraction of their quark counterparts. By adopting Gaussian models for the gluon TMDs, which fulfill without saturating everywhere their positivity bounds, we provide numerical predictions for the transverse single-spin asymmetries.
These observables could be measured at LHCSpin, the fixed target experiment planned at the LHC.
\end{abstract}

\date{\today}
\maketitle
\section{Introduction}

It is well-known that bound states of heavy quarks (quarkonia) produced in proton-proton collisions can be considered as direct probes of the gluon content of the proton, providing detailed information about gluon momentum distributions and in particular their transverse momentum dependence. From the theoretical point of view, quarkonia with even charge conjugation ($C=+1$) are produced by the fusion of two gluons in a $2 \to 1$ partonic reaction at leading order in the strong coupling constant $\alpha_s$, with no additional gluon emission in the final state. Thus, in analogy to the Drell-Yan processes, the kinematics are very simple, with gluon momentum fractions directly related to the rapidity of the observed quarkonium state ${\cal Q}$. Furthermore, the charm and bottom masses are large enough to justify the use of perturbative QCD even when the transverse momentum $q_\sT$ of the quarkonium state is small, namely $q^2_\sT \ll M^2_{\cal Q}$. In this kinematic region transverse momentum dependent (TMD) factorization is expected to be applicable. Here we mainly focus on scalar and pseudoscalar $C$-even quarkonia, {\it i.e.}\ states with definite total angular momentum $J$, parity $P$ and charge conjugation $J^{PC} = 0^{\pm +}$. Namely, using the alternative spectroscopic notation $^{2S+1}L_J$, with $S$ being the spin and $L$ the orbital angular momentum, we consider the $^1S_0$ states $\eta_c,\eta_b$ and  the $^3 P_0$ states $\chi_{c0}, \chi_{b0}$. The $^3 P_2$ ($2^{++}$) states $\chi_{c2}$ and $\chi_{b2}$  are studied as well, whereas the $\chi_{c1}$ and $\chi_{b1}$ states would require a different treatment because they suffer from the same problem as other vector states, such as the $J/\psi$ meson. Indeed, due to the Landau-Yang theorem, their production from two on-shell gluons requires the emission of an additional gluon. 

As pointed out in Refs.~\cite{Boer:2012bt,Brodsky:2012vg}, another advantage of dealing with $C$-even quarkonia is that they suffer neither from large QCD corrections nor from the many open theoretical issues affecting the predictions for $J/\psi$ and $\Upsilon$ production rates and polarization~\cite{Brambilla:2010cs,Andronic:2015wma,Lansberg:2019adr}. 
 The latter statement can be understood by employing the effective field theory approach of nonrelativistic QCD (NRQCD)~\cite{Bodwin:1994jh}, according to which quarkonium production in proton-proton collisions is described in terms of a double power series expansion in $\alpha_s$ and the relative velocity $v$ of the heavy quark-antiquark pair in the quarkonium rest frame, with $v \ll 1$. The magnitude of the velocity is given by $v^2 \simeq 0.3$ for charmonium and $v^2 \simeq 0.1$ for bottomonium. Within this framework, a heavy quark-antiquark pair can be produced at short distances not just as a color-singlet, but also in a color-octet configuration, which subsequently evolves into a physical quarkonium state by radiating soft gluons. The hadronization of the pair is encoded in universal long-distance matrix elements (LDMEs), which are expected to scale with a definite power of $v$. These matrix elements are not calculable perturbatively and have to be extracted from data. For $S$-wave quarkonia, in the limit $v\to 0$ the heavy quark-antiquark pair is produced directly with the same quantum numbers of the observed bound state and the traditional color-singlet model (CSM)~\cite{Kuhn:1979bb,Guberina:1980dc,Baier:1983va} is recovered. While the CSM fails to describe the large transverse momentum spectra of the $C$-odd vector states $J/\psi$, $\psi(2S)$ and $\Upsilon$, that should not be the case for $C$-even states, for which NRQCD shows that color-octet contributions are (at least) order $v^2$ suppressed with respect to the color-singlet ones~\cite{Bodwin:1994jh}. This is confirmed by the study of the low transverse momentum part of the spectrum of $\chi_{c1,2}$~\cite{Bodwin:2005hm}. More recently, it has been found that the CSM provides an excellent description of the LHCb data on inclusive $\eta_c$ production~\cite{Butenschoen:2014dra}, probed through the $p \overline p$ decay channel~\cite{LHCb:2014oii}. Furthermore, according to Ref.~\cite{Brodsky:2012vg}, color-octet contributions can certainly be neglected for $C$-even bottomonium, in agreement with the analysis on $\eta_b$ mesons of Ref.~\cite{Maltoni:2004hv}. 

Based on the above considerations, in this paper we employ the CSM in combination with TMD factorization to study the effects of gluon distributions on $0^{\pm +}$ and $2^{++}$ quarkonia produced in proton-proton collisions, with one or both protons being polarized. Following the same approach, in Ref.~\cite{Boer:2012bt} the distribution of linearly polarized gluons inside an unpolarized proton, named $h_1^{\perp\, g}$, has been investigated. This function corresponds to an interference between $+1$ and $-1$ helicity gluon states that would be suppressed without transverse momentum. It has been shown that it modifies the unpolarized cross sections for the production of scalar and pseudoscalar scalar quarkonia in different ways, while its effect on higher angular momentum states is strongly suppressed.  
Similarly, here we show how single and double spin asymmetries arise from other helicity-flip distributions of linearly polarized gluons inside transversely polarized protons  ($h_1^{g}$, $ h_{1T}^{\perp\,g}$) and longitudinally polarized protons ($h_{1L}^{\perp\,g}$). Furthermore, the Sivers function~\cite{Sivers:1989cc,Boer:2015vso} needs to be taken into account, which describes the transverse momentum distribution of unpolarized quarks and gluons inside a transversely polarized proton, where the transverse momentum forms a $\sin\phi$ distribution around the transverse spin direction. Using the NRQCD approach together with TMD factorization, it has indeed been shown that the gluon Sivers function generates a single spin asymmetry only in the CSM in proton-proton collisions, and only in the color-octet model in lepton-proton collisions~\cite{Yuan:2008vn}.  

Because of their gauge link dependence, TMDs are not universal. 
For the processes under study, gauge links are exclusively past pointing, $[-,-]$, as for Higgs~\cite{Sun:2012vc,Boer:2011kf,Echevarria:2015uaa,Gutierrez-Reyes:2019rug} or photon pair production~\cite{Qiu:2011ai} in hadronic collisions. About quarkonium final states at the LHC, the same gauge-link structure holds for $J/\psi$-photon~\cite{denDunnen:2014kjo} and double-$J/\psi$~\cite{Lansberg:2017dzg, Serri:2024qne} production as well, assuming the dominance of the color-singlet quarkonium formation mechanism. A global analysis of gluon TMDs from the above reactions would have the advantage of mapping out their scale dependence. On the other hand, their universality properties can be tested by relating the $[-,-]$ gluon TMDs to the $[+,+]$ ones, with two future-pointing gauge links, contributing for instance to dijet, open heavy-quark pair~\cite{Boer:2016fqd}, inclusive $J/\psi$~\cite{Mukherjee:2016qxa,Bacchetta:2018ivt}, $J/\psi$-jet~\cite{DAlesio:2019qpk,Kishore:2019fzb} and $J/\psi$-photon~\cite{Chakrabarti:2022rjr} production in electron-proton collisions, which are in principle accessible at the future Electron-Ion Collider (EIC). The $[-,-]$ gluon TMDs investigated in this paper correspond to the Weiszäcker-Williams (WW) distributions at small $x$. It turns out that, unlike the $[+,-]$ or dipole ones, which have one future and one past pointing gauge link, the WW gluon TMDs for a transversely polarized proton are suppressed with respect to the unpolarized gluon distribution by a factor of $x$~\cite{Boer:2015pni}. This implies that the transverse spin asymmetries under study will become suppressed in the small-$x$ limit. 

The proposed measurements require the observation of $C$-even quarkonium states with small transverse momentum, resulting from the transverse momenta of the partons initiating the $2\to1$ reactions. At collider facilities like the LHC, forward detectors such as LHCb are needed, together with powerful particle identification for a complete study of the different quarkonium states through their decay channels. Moreover, the possibility of having a polarized gas target in front of the LHCb spectrometer, the so-called LHCSpin experiment~\cite{Aidala:2019pit}, will offer the unique opportunity to probe polarized gluon TMD through the analysis of single-spin asymmetries (SSAs). 

The remainder of the paper is organized as follows. In Section~\ref{sec:TMD} we recall the definition of the gluon correlator and the leading twist TMD distributions in terms of QCD operators. Details of the
calculation of the cross sections of interest, together with the analytic results for the azimuthal modulations, are presented in Section~\ref{sec:formalism}. The Fourier transforms in transverse position space of the expressions for the gluon correlator and the convolutions of TMDs contributing to the azimuthal asymmetries are listed in Appendix~\ref{app: convolution bT-space}.
Our numerical estimates of the transverse single-spin asymmetries obtained by adopting Gaussian models for the gluon densities can be found in Section~\ref{sec: numerical results}. Finally, Section~\ref{sec:conclusions} contains our summary and conclusions.

\section{Transverse momentum dependent gluon distributions}
\label{sec:TMD}
\begin{figure}[t]
    \centering
    \includegraphics[width=0.38\textwidth]{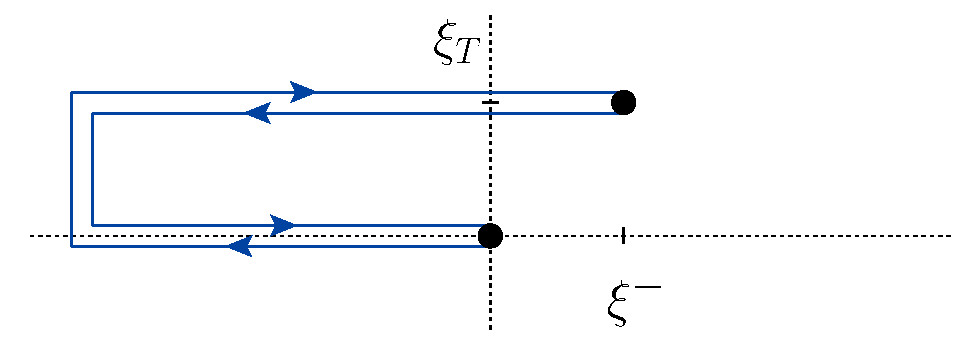}
    \caption{Illustration of the $[--]$ gauge link structure. The horizontal axis corresponds to the light-cone direction $n_-$, while the vertical one represents the two transverse directions. The two dots denote the points 0 and $\xi$.}
    \label{fig:links}
\end{figure}

Transverse momentum dependent gluon distribution functions in a spin-1/2 hadron are defined through a matrix element of a correlator of the gluon field strengths $F^{\mu \nu}(0)$ and $F^{\nu \sigma}(\xi)$, evaluated at fixed light-front (LF) time $\xi^+ = \xi{\cdot}n = 0$, where $n$ is a lightlike vector conjugate to the four-momentum $P$ of the parent hadron. Decomposing the gluon momentum as
$p = x\,P + p_\sT + p^- n$, the correlator is given by~\cite{Mulders:2000sh,Meissner:2007rx,Boer:2016xqr}
\begin{equation}
\label{GluonCorr}
\Gamma_g^{[U,U^\prime]\mu\nu}(x,\bm p_\sT )
= \frac{n_\rho\,n_\sigma}{(p{\cdot}n)^2}
{\int}\frac{\d(\xi{\cdot}P)\,\d^2\xi_\sT}{(2\pi)^3}\
e^{ip\cdot\xi}\,
\langle P, S|\,\tr\big[\,F^{\mu\rho}(0)\, U_{[0,\xi]}
F^{\nu\sigma}(\xi)\,U^\prime_{[\xi,0]}\,\big]
\,|P, S \rangle\,\big\rfloor_{\text{LF}}\,,
\end{equation}
where $S$ is the hadron spin vector, while $U_{[0,\xi]}$ and $U^\prime_{[0,\xi]}$ are process-dependent gauge links, which make the correlator gauge invariant. For the process under study, $U_{[0,\xi]}$ is a  staple-like Wilson line running from 0 to $\xi$, namely  $U^{[-]}_{[0,\xi]} = U^{[n]}_{[0,-\infty]} U^\sT_{[0_\sT, \xi_\sT]} U^{[n]}_{[-\infty, \xi]}$, while $U^\prime_{[\xi,0]}$ runs from $\xi$ to 0 and is given by $U^{[-]}_{[\xi,0]} = U^{[-]\dagger}_{[0,\xi]}$, as illustrated in Fig.~\ref{fig:links}. Henceforth, the explicit dependence on the gauge links will be omitted.

According to the hadron spin, the correlator can be split into three parts: the unpolarized ($U$), the longitudinal polarized ($L$) and the transversely polarized ($T$) components,  
\begin{align}
\Gamma_{g}^{\mu\nu}(x,\bm p_\sT ) =  \Gamma_{g\, U}^{\mu\nu}(x,\bm p_\sT ) + \Gamma_{g\, L}^{\mu\nu}(x,\bm p_\sT ) + \Gamma_{g\, T}^{\mu\nu}(x,\bm p_\sT )  \,.
\label{eq:Gamma-parts}
\end{align}
 At leading twist, the correlator for an unpolarized hadron can be parameterized in terms of two gluon TMDs as follows:
\begin{align}
\Gamma_{g\, U}^{\mu\nu}(x,\bm p_\sT ) & =  \frac{1}{2x}\,\bigg \{-g_\sT^{\mu\nu}\,f_1^g (x,\bm p_\sT^2) +\bigg(\frac{p_\sT^\mu p_\sT^\nu}{M_h^2}\,
    {+}\,g_\sT^{\mu\nu}\frac{\bm p_\sT^2}{2M_h^2}\bigg) \,h_1^{\perp\,g} (x,\bm p_\sT^2) \bigg \} \,,
\label{eq:Gamma-U}
\end{align}
where $p_{\sT}^2 = -\bm p_{\sT}^2$, $M_h$ is the hadron mass and the symmetric transverse projector $g^{\mu\nu}_{\sT}$ is defined as $g^{\mu\nu}_{\sT} = g^{\mu\nu} - P^{\mu}n^{\nu}/P{\cdot}n-n^{\mu}P^{\nu}/P{\cdot}n$. Furthermore, in Eq.~\eqref{eq:Gamma-U} $f_1^g$ and $h_1^{\perp\,g}$ are the T-even unpolarized and linearly polarized gluon distributions, respectively. 
In order to write the other two correlators in a convenient form, we define the longitudinal and transverse components of the hadron spin through the Sudakov decomposition
\begin{align}
S^\mu = \frac{S_\sL}{M_h}\left ( P^\mu - \frac{M_h^2}{P\cdot n}\,n^\mu \right ) + S_\sT^\mu\,,
\label{eq:Gamma-L}
\end{align}
with $S_\sL^2 + \bm S_\sT^2=1$. Hence, the correlator for a longitudinally polarized hadron can be written as
\begin{align}
\Gamma_{g\, L}^{\mu\nu}(x,\bm p_\sT ) & = \frac{1}{2x}\, S_\sL\,
    \left \{ i \epsilon^{\mu\nu}_\sT \,
    g_{1 \sL}^g (x, \bm p_\sT^2) \, + \,
    \frac{\epsilon_\sT^{p_\sT \{ \mu} p_\sT^{\nu \}}}{M_h^2} \, h_{1\sL}^{\perp\,g}(x,\bm p_\sT^2) \right \}\,,  
\label{eq:Gamma-T}
\end{align}
where we have introduced the antisymmetric transverse projector $\epsilon_\sT^{\mu\nu} = \epsilon_\sT^{\alpha\beta\mu\nu} P_\alpha n_\beta/P{\cdot}n$, with $\epsilon_\sT^{1 2} = +1$, as well as the notations 
$\epsilon^{a b}_{\sT} \equiv \epsilon^{\alpha\beta}_{\sT} a_{\alpha} \,b_{\beta}$ and
$a^{\{\mu}b^{\nu \}} \equiv a^\mu b^\nu + a^\nu b^\mu$. In Eq.~\eqref{eq:Gamma-T} $g^g_{1\sL}$ is the T-even helicity distribution, whereas $h^{\perp\,g}_{1\sL}$ is the T-odd distribution of linearly polarized gluons inside a longitudinally polarized hadron. Finally, for a transversely polarized hadron
\begin{align}    
    \Gamma_{g\, T}^{\mu\nu}(x,\bm p_\sT ) & = \frac{1}{2x}\,\bigg \{g^{\mu\nu}_\sT\,
    \frac{ \epsilon^{p_\sT S_\sT}_\sT}{M_h}\, f_{1\sT}^{\perp\,g}(x, \bm p_\sT^2)\, +\, i \epsilon_\sT^{\mu\nu}\,
    \frac{p_\sT \cdot S_\sT}{M_h}\,g_{1\sT}^{\perp\,g}(x, \bm p_\sT^2)\,  -\,\frac{\epsilon_\sT^{p_\sT \{ \mu}S_\sT^{\nu \}}\,+\,\epsilon_\sT^{S_\sT\{ \mu } p_\sT^{\nu \}}}{4M_h} \, h_{1}^{g}(x, \bm p_\sT^2) \nonumber\\
 & \phantom{=} + \,  \frac{4\,(p_\sT\cdot S_\sT)\,\epsilon_\sT^{p_\sT \{ \mu}p_\sT^{\nu \}} + \bm  p_\sT^2 \left [ \epsilon_\sT^{p_\sT \{ \mu}S_\sT^{\nu \}}+ 
      \epsilon_\sT^{S_\sT\{ \mu } p_\sT^{\nu \} }\right ] }{8M_h^3}\, h_{1 \sT}^{\perp\,g}(x, \bm p_\sT^2)\,\bigg \}\,.
\end{align}
Note that the symmetric part of the correlator, $(\Gamma_\sT^{\mu\nu} + \Gamma^{\nu\mu}_{\sT})/2$, is parametrized in terms of three T-odd distributions, namely the Sivers function $f_{1\sT}^{\perp\,g}$ and the two helicity-flip distributions $h_1^g$ and $h_{1\sT}^{\perp\,g}$. On the other hand, the TMD distribution $g_{1\sT}^{\perp\,g}$ is T-even.

Because of the lack of data on processes that could be used for extractions, our current knowledge of gluon TMDs is still very limited. Recently a small set of LHCb data on double-$J/\psi$ production at 13 TeV~\cite{LHCb:2016wuo} has been used to perform a first fit of $f_1^g$, assuming a simple Gaussian dependence on the parton transverse momentum~\cite{Lansberg:2017dzg}. 
Theoretical computations of gluon TMDs within the color glass condensate (CGC) framework \cite{Metz:2011wb,Dominguez:2011br,Boer:2016fqd}  and spectator models~\cite{Bacchetta:2020vty,Chakrabarti:2023djs,Yu:2024mxo,Bacchetta:2024fci} have also been carried out.

\section{Azimuthal modulations at small transverse momentum}
\label{sec:formalism}
\begin{figure}[t]
    \centering
    \includegraphics[width=0.475\linewidth, keepaspectratio]{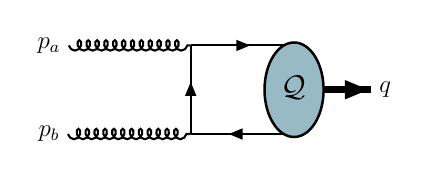}
    \caption{Leading order diagram for the process 
    $gg \to {\cal Q}$, where ${\cal Q}$ is a heavy quark-antiquark bound state with quantum numbers $^{2S+1}L^{(1)}_J$. 
   The crossed diagram, in which the directions of the arrows in the fermionic lines are reversed, is not shown.}
    \label{fig: gg leading-order}
\end{figure}

We consider the inclusive scattering process 
\begin{equation}
p(P_A, S_A)\,{+}\,p(P_B, S_B)\,\to\, Q \overline Q [^{2S+1}L_J^{(1)}](q)\, {+}  \,X \, ,
\label{eq:process}
\end{equation}
where the two colliding protons or, more in general, two spin-1/2 hadrons have four-momenta $P_A$ and $P_B$ and spin vectors $S_A$ and $S_B$, such that $S_A^2= S_B^2= -1$ and $S_A\cdot P_A = S_B\cdot P_B =0$.
We assume that a heavy quark-antiquark pair ($Q\overline Q$) is produced in an intermediate Fock state with four-momentum $q$, spin $S$, orbital angular momentum $L$, total angular momentum $J$ and in a colorless configuration, specified by the superscript $(1)$. The squared invariant mass of the resonance is $M^2 = q^2$, with $M$ twice the heavy quark mass up to small relativistic corrections.
According to the CSM, these quantum numbers match the ones of the outgoing observed quarkonium. 
Hence, smearing effects in the hadron formation process encoded in the so-called TMD shape functions~\cite{Echevarria:2019ynx,Fleming:2019pzj} are expected to be suppressed~\cite{Boer:2023zit,Sun:2012vc} and will therefore be neglected.
At the lowest order in perturbative QCD, one only has to consider the gluon-gluon fusion process
\begin{equation}
g(p_a)\,{+}\,g(p_b)\,\to\, Q \overline Q [^{2S+1}L_J^{(1)}] (q)\, ,
\label{eq:subproc}
\end{equation}
which is described by the Feynman diagram depicted in~\cref{fig: gg leading-order}.

In the kinematic region where the transverse momentum $q_\sT$ of the produced quarkonium state is much smaller than its invariant mass, namely $q_\sT \ll M$, TMD factorization is expected to be applicable and the cross section for the process in Eq.~\eqref{eq:process} can be written as 
\begin{equation}
\d\sigma
= \frac{1}{2 s}\,\frac{\d^3 \bm q}{(2\pi)^3\,2 q^0} 
{\int} \d x_a \,\d x_b \,\d^2\bm p_{a\sT} \,\d^2\bm p_{b\sT}\,(2\pi)^4
\delta^4(p_a{+} p_b {-} q)\,
\overline{\sum_{\rm colors}} \,\Gamma_g^{\mu\nu}(x_a {,}\bm p_{a \sT})\, \Gamma_g^{\rho\sigma}(x_b {,}\bm p_{b \sT})\, {\cal A}_{\mu\rho} \,({\cal A}_{\nu\sigma})^* 
\label{eq:CrossSec}
\end{equation}
where $s = (P_A + P_B)^2$ is the total energy squared in the hadronic center-of-mass frame, $\Gamma_g$ is the gluon correlator parametrized in Eqs.~\eqref{eq:Gamma-parts}-\eqref{eq:Gamma-T} and ${\cal A}$ is the scattering amplitude for the specific partonic process $gg\to {\cal Q}$. 
We note that the partonic momenta fulfill the relation $p_a^-=p_b^+ =0$ in our calculation. 
Moreover, for the proton with momentum $P_B$ the role of the forward and backward light-cone directions is exchanged as compared to the other proton with momentum $P_A$, hence in Eqs.~\eqref{eq:Gamma-L}-\eqref{eq:Gamma-T} the epsilon tensor should be taken with opposite sign: $\epsilon_\sT^{\mu\nu} \to -\epsilon_\sT^{\mu\nu}$.

It turns out that the only nonzero scattering amplitudes correspond to the $^1S_0$ $(\eta_Q)$ and $^3P_{0,2}$ $(\chi_{Q{0,2}})$ states, where $Q = c, b$~\cite{Boer:2012bt}:
\begin{align}
	&\mathcal{A}^{\mu\nu} [^1S_0^{(1)} ] (p_a,p_b;q) = \, 2i\frac{\delta^{ab}}{\sqrt{N_c}} \frac{g^2_s}{\sqrt{\pi M^5}} R_0(0) \epsilon^{\mu\nu\rho\sigma}p_{a\rho}p_{b\sigma}\,,\nonumber\\
&\mathcal{A}^{\mu\nu}[^3P^{(1)}_0](p_a,q) =\, -2i\frac{\delta^{ab}}{\sqrt{N_c}} \frac{g^2_s}{\sqrt{\pi M^3}} R'_1(0) \bigg[-3g^{\mu\nu}+\frac{2}{M^2}q^{\mu}p^{\nu}_a\bigg]\,,\\
	&\mathcal{A}^{\mu\nu}[^3P^{(1)}_2](p_a,q)=\, -2i\frac{\delta^{ab}}{\sqrt{N_c}}\sqrt{\frac{3}{\pi M^3}} g^2_s R'_1(0) \varepsilon_{J_z}^{\rho\sigma}(q) \bigg[\frac{4}{M^2}g^{\mu\nu}p_{a\rho}p_{a\sigma}-g^{\mu}_{	\rho}g^{\nu}_{\sigma}-g^{\nu}_{\rho}g^{\mu}_{\sigma}\bigg]\,,\nonumber
\end{align}
with $N_c$ being the number of colors, $g_s$ the QCD coupling and $\epsilon_{J_z}$ the polarization vector of the $J=2$ bound state. Furthermore,  $R_L(r)$ is the radial wave function of the quarkonium state with orbital angular momentum $L$ and $R^\prime_L(r)$ its first derivative, in terms of which the NRQCD LDMEs are given by 
\begin{equation}
\begin{aligned}
	\bra{0}\mathcal{O}^{\eta_Q}_1(^1S_0)\ket{0} & = \frac{N_c}{2\pi} |R_0(0)|^2[1+\mathcal{O}(v^4)]\,, \\
	\bra{0}\mathcal{O}^{\chi_{QJ}}_1(^3P_J)\ket{0} & = \frac{3N_c}{2\pi}|R'_1(0)|^2[1+\mathcal{O}(v^2)]\,,
 \label{eq:LDMEs}
\end{aligned}
\end{equation}
with $J = 0,1,2$ and $v$ being the relative velocity of the heavy
quark-antiquark pair in the quarkonium rest frame. We point out that for these quarkonium states TMD factorization has been proven at one loop level~\cite{Ma:2012hh,Ma:2014oha}.

The differential cross section in~\cref{eq:CrossSec} for the production of a generic $C=+1$ quarkonium state, in a frame where its azimuthal angle is zero, namely $\phi_\sT = 0$, can be written as
\begin{align}
    \frac{\d \sigma[{\cal Q}]}{\d y\,\d^2 \bm q_\sT} & = 
    F^{\cal Q}_{UU} + F^{\cal Q}_{UL}\, S_{B \sL} + F^{\cal Q}_{LU}\, S_{A \sL} + F_{UT}^{{\cal Q}, \sin\phi_{S_B}}\, |\bm S_{B \sT}|\sin\phi_{S_B} + F_{TU}^{{\cal Q}, \sin\phi_{S_A}}\, |\bm S_{A \sT}| \sin\phi_{S_A} \nn \\ & \phantom{=}
    + F^{\cal Q}_{LL}\, S_{A \sL}\, S_{B \sL} + F_{LT}^{{\cal Q}, \cos\phi_{S_B}}\, S_{A \sL}\, |\bm S_{B \sT}|\cos\phi_{S_B} + F_{TL}^{{\cal Q}, \cos\phi_{S_A}}\, |\bm S_{A \sT}|\, S_{B \sL} \cos\phi_{S_A} \nn \\ & \phantom{=}
    + |\bm S_{A \sT}|\, |\bm S_{B \sT}|  \left(F_{TT}^{{\cal Q}, \cos(\phi_{S_A}- \phi_{S_B})
    }\, \cos(\phi_{S_A} - \phi_{S_B})
     +F_{TT}^{{\cal Q}, \cos(\phi_{S_A} + \phi_{S_B})
    }\, \cos(\phi_{S_A} + \phi_{S_B}) \right)  \, ,
\label{eq: diff cs full parameters}
\end{align}
with $y$ and $\bm q_\sT$ being the rapidity and the transverse momentum of the outgoing quarkonium, respectively. Furthermore, $\phi_{S_A}$ {($\phi_{S_B}$) is the azimuthal angle of the spin vector $S_A$ ($S_B$). The subscripts of the structure functions $F^{\cal Q}$ refer to the polarization of the incoming protons.  Each structure function in \cref{eq: diff cs full parameters} can be factorized in a hard part $H^{\cal Q}$, which is calculable as a perturbative expansion in $\alpha_s$, 
\begin{equation}
    H^{\cal Q} = \sum_{n = 0}^\infty \left( \frac{\alpha_s}{\pi} \right)^n H^{\cal Q}_n \, ,
\label{eq: hard term expansion}
\end{equation}
and a nonpertubative part, given by one or more convolutions of gluon TMDs multiplied by one of the LDMEs in Eq.~\eqref{eq:LDMEs}. In momentum space, these convolutions are defined as
\begin{align}
    {\cal C} [w\, f^g_1\, f^g_2] & 
    \equiv {\cal C} [w(\bm p_{a\sT}, \bm p_{b\sT})\, f^g_1(x_a,\bm p_{a\sT})\, f^g_2(x_b,\bm p_{b\sT})] \nn \\ & 
    = \int \d^2\bm p_{a\sT} \,\d^2\bm p_{b\sT}\, w(\bm p_{a\sT}, \bm p_{b\sT})\, f^g_1(x_a,\bm p_{a\sT})\, f^g_2(x_b,\bm p_{b\sT})\, \delta^{2}(\bm p_{a\sT} + \bm p_{b\sT} - \bm q_\sT)\, ,
\end{align}
where $f_i$, with $i=1,2$, are the gluon TMDs and $w(\bm p_{a\sT}, \bm p_{b\sT})$ is a proper weight function that depends on the particular gluon distributions involved. Neglecting terms suppressed by powers of $\bm q_\sT/M$, for the process under study the light-cone momentum fractions are given by
\begin{equation}
    x_a = \frac{M}{\sqrt s} \e^y \, , \qquad  
    x_b = \frac{M}{\sqrt s} \e^{-y} \, .
\end{equation}

In the following we provide the explicit expressions of the structure functions in \cref{eq: diff cs full parameters} at the order $\alpha_s^2$, as obtained from the evaluation of diagrams like the one in \cref{fig: gg leading-order}.
We note that higher-order corrections will not modify the TMD convolutions involved, but more terms will have to be included in the $\alpha_s$-expansion in \cref{eq: hard term expansion}. 
In particular, the fully unpolarized structure functions $F_{UU}^Q$ for $J=0$ and $J=2$ quarkonia read
\begin{equation}
\begin{aligned}
    F_{UU}^{\eta_Q} & = H^{\eta_Q}\, \Big(
        {\cal C}\big[f_1^g f_1^g\big] - {\cal C}\big[w_{\scriptscriptstyle UU}\, h_1^{\perp\, g} h_1^{\perp\, g}\big] \Big)\, \langle 0\vert {\cal{O}}^{\eta_Q}_1 (^1 S_0)\vert 0\rangle \, , \\
    F_{UU}^{\chi_{Q0}} & = H^{\chi_{Q0}}\, \Big(
        {\cal C}\big[f_1^g f_1^g\big] + {\cal C}\big[w_{\scriptscriptstyle UU}\, h_1^{\perp\, g} h_1^{\perp\, g}\big] \Big)\, \langle 0\vert {\cal{O}}^{\chi_{Q0}}_1 (^3 P_0)\vert 0\rangle \, , \\
    F_{UU}^{\chi_{Q2}} & = H^{\chi_{Q2}}\, 
        {\cal C}\big[f_1^g f_1^g\big] \, \langle 0\vert {\cal{O}}^{\chi_{Q2}}_1 (^3 P_2)\vert 0\rangle 
    \label{eq: FUU}\, ,
\end{aligned}
\end{equation}
with
\begin{equation}
\begin{aligned}
    H^{\eta_Q}_{0} & = \frac{2\, \pi^3\alpha_s^2}{9 M^3 s}\, , \\
    H^{\chi_{Q0}}_{0} & = \frac{8\, \pi^3\alpha_s^2}{3 M^5 s} \, , \\
    H^{\chi_{Q2}}_{0} & = \frac{32\, \pi^3\alpha_s^2}{9 M^5 s} \, ,
\end{aligned}
\end{equation}
and the weight function being given by
\begin{equation}
    w_{\scriptscriptstyle UU} = \frac{1}{4 M_p^4} \big[ 2\, (\bm p_{a\sT} \cdot \bm p_{b \sT})^2 - \bm p_{a\sT}^2 \bm p_{b\sT}^2 \big] = \frac{\bm p_{a\sT}^2 \bm p_{b\sT}^2}{4 M_p^4} \cos \big[ 2(\phi_a - \phi_b)\big] \, ,
\end{equation}
with $\phi_a$ and $\phi_b$ being the azimuthal angles of $\bm p_{a\sT}$ and $\bm p_{b\sT}$, respectively. The results in Eq.~\eqref{eq: FUU} are in agreement with Ref.~\cite{Boer:2012bt} and, upon integration over $\bm q_\sT$, with Ref.~\cite{Baier:1983va}.
Hence, we confirm all the features discussed in Ref.~\cite{Boer:2012bt}, and in particular the sign difference in the term ${\cal C}\big[w_{\scriptscriptstyle UU}\, h_1^{\perp\, g} h_1^{\perp\, g}\big]$ for opposite parities of the $J=0$ quarkonium states, and its absence for $\chi_{Q2}$. 
Indeed, a nonzero contribution from linearly polarized gluons for $J=2$ states would require a four-unit helicity flip, which is strongly suppressed. Hence, at leading order in $v^2$, the ratio of the cross sections for $\chi_{Q0}$ and $\chi_{Q2}$ can be used as a direct probe of the quantity ${\cal C}\big[w_{\scriptscriptstyle UU}\, h_1^{\perp\, g} h_1^{\perp\, g}\big]/ {\cal C}\big[f_1^g f_1^g\big]$. Note that both the uncertainties from the unpolarized gluon TMD and the hadronic matrix elements cancel out in this ratio, see Eq.~\eqref{eq:LDMEs}.

To the best of our knowledge, the explicit expressions of the other structure functions in \cref{eq: diff cs full parameters}, depending on single and double polarization effects of the initial protons, are presented here for the first time. Due to parity conservation, for the single-longitudinally polarized contributions one has
\begin{equation}
    \begin{aligned}
        F_{UL}^{\eta_Q} & = F_{LU}^{\eta_Q} = 0 \, , \\
        F_{UL}^{\chi_{Q0}} & = F_{LU}^{\chi_{Q0}} = 0 \, , \\
        F_{UL}^{\chi_{Q2}} & = F_{LU}^{\chi_{Q2}} = 0
        \label{eq: FUL}\, .
    \end{aligned}
\end{equation}
Therefore, only the transverse polarization is relevant in the single polarization case. We find
\begin{equation}
    \begin{aligned}
        F_{UT}^{\eta_Q, \sin \phi_{S_B}} & = H^{\eta_Q} \Big(
            - {\cal C}\big[w_{\scriptscriptstyle UT}^f\, f_1^g f_{1\sT}^{\perp\, g} \big] + {\cal C}\big[w_{\scriptscriptstyle UT}^{h}\, h_1^{\perp\, g} h_1^{g}\big] - {\cal C}\big[w_{\scriptscriptstyle UT}^{h^\perp}\, h_1^{\perp\, g} h_{1\sT}^{\perp\, g}\big] \Big)\, \langle 0\vert {\cal{O}}^{\eta_Q}_1 (^1 S_0)\vert 0\rangle \, , \\
        F_{UT}^{\chi_{Q0}, \sin \phi_{S_B}} & = H^{\chi_{Q0}} \Big(
            - {\cal C}\big[w_{\scriptscriptstyle UT}^f\, f_1^g f_{1\sT}^{\perp\, g} \big] - {\cal C}\big[w_{\scriptscriptstyle UT}^{h}\, h_1^{\perp\, g} h_1^{g}\big] + {\cal C}\big[w_{\scriptscriptstyle UT}^{h^\perp}\, h_1^{\perp\, g} h_{1\sT}^{\perp\, g}\big] \Big)\, \langle 0\vert {\cal{O}}^{\chi_{Q0}}_1 (^3 P_0)\vert 0\rangle \, , \\
        F_{UT}^{\chi_{Q2}, \sin \phi_{S_B}} & = - H^{\chi_{Q2}} \, 
             {\cal C}\big[w_{\scriptscriptstyle UT}^f\, f_1^g f_{1\sT}^{\perp\, g} \big] \, \langle 0\vert {\cal{O}}^{\chi_{Q2}}_1 (^3 P_2)\vert 0\rangle \, ,
    \end{aligned}
\label{eq:F_UT}
\end{equation}
with
\begin{equation}
    \begin{aligned}
        w_{\scriptscriptstyle UT}^f & = 
        \frac{|\bm p_{b\sT}|}{M_p} \cos \phi_b \, , \\
        w_{\scriptscriptstyle UT}^h & = 
        \frac{\bm p_{a\sT}^2\, |\bm p_{b\sT}|}{4 M_p^3} \cos( \phi_b - 2 \phi_a) \, , \\
        w_{\scriptscriptstyle UT}^{h^\perp} & =  
        \frac{\bm p_{a\sT}^2\, |\bm p_{b\sT}|^3}{8 M_p^5} \cos(3 \phi_b - 2 \phi_a) \, ;
    \end{aligned}
\label{eq: wUT}
\end{equation}
and
\begin{equation}
\begin{aligned}
    F_{TU}^{\eta_Q, \sin \phi_{S_A}} & = H^{\eta_Q} \Big(
        {\cal C}\big[w_{\scriptscriptstyle TU}^f\, f_1^g f_{1\sT}^{\perp\, g} \big] - {\cal C}\big[w_{\scriptscriptstyle TU}^{h}\, h_1^{\perp\, g} h_1^{g}\big] + {\cal C}\big[w_{\scriptscriptstyle TU}^{h^\perp}\, h_1^{\perp\, g} h_{1\sT}^{\perp\, g}\big] \Big)\, \langle 0\vert {\cal{O}}^{\eta_Q}_1 (^1 S_0)\vert 0\rangle \, , \\
    F_{TU}^{\chi_{Q0}, \sin \phi_{S_A}} & = H^{\chi_{Q0}} \Big(
        {\cal C}\big[w_{\scriptscriptstyle TU}^f\, f_1^g f_{1\sT}^{\perp\, g} \big] + {\cal C}\big[w_{\scriptscriptstyle TU}^{h}\, h_1^{\perp\, g} h_1^{g}\big] - {\cal C}\big[w_{\scriptscriptstyle TU}^{h^\perp}\, h_1^{\perp\, g} h_{1\sT}^{\perp\, g}\big] \Big)\, \langle 0\vert {\cal{O}}^{\chi_{Q0}}_1 (^3 P_0)\vert 0\rangle \, , \\
    F_{TU}^{\chi_{Q2}, \sin \phi_{S_A}} & = H^{\chi_{Q2}} \, 
        {\cal C}\big[w_{\scriptscriptstyle TU}^f\, f_1^g f_{1\sT}^{\perp\, g} \big] \, \langle 0\vert {\cal{O}}^{\chi_{Q2}}_1 (^3 P_2)\vert 0\rangle \, ,
        \label{eq:F_TU}
\end{aligned}
\end{equation}
where the weight functions in Eq.~\eqref{eq:F_TU} can be obtained from the ones in Eq.~\eqref{eq: wUT} with the replacements $U\leftrightarrow T$ and $a \leftrightarrow b$.
Analogously to $F^{\cal Q}_{UU}$, the structure functions in Eqs.~\eqref{eq:F_UT} and \eqref{eq:F_TU} for the $\chi_{Q2}$ mesons receive a contribution solely from unpolarized gluon distributions, namely $f_1^g$ and $f_{1\sT}^{\perp\,g}$. Moreover, the convolutions involving distributions of linearly polarized gluons, ${\cal C}\big[w_{\scriptscriptstyle UT}\, h_1^{\perp\, g} h_1^g\big]$ and ${\cal C}\big[w_{\scriptscriptstyle UT}\, h_1^{\perp\, g} h_{1\sT}^{\perp\, g}\big]$, enter the structure functions for $J=0$ states with a sign
depending on the quarkonium parity. 
As will be further discussed in \cref{sec: numerical results}, in principle a combined analysis of these observables could allow to probe the gluon TMDs $f_{1\sT}^{\perp\,g}$, $h_1^g$ and $h_{1\sT}^{\perp\, g}$.

The double-longitudinally polarized structure functions read
\begin{equation}
    \begin{aligned}
        F_{LL}^{\eta_Q} & = H^{\eta_Q}\, \Big(
            {\cal C}\big[g_{1\sL}^g g_{1\sL}^g\big] + {\cal C}\big[w_{\scriptscriptstyle LL}\, h_{1\sL}^{\perp\,g} h_{1\sL}^{\perp\,g}\big] \Big)\, \langle 0\vert {\cal{O}}^{\eta_Q}_1 (^1 S_0)\vert 0\rangle \, , \\
        F_{LL}^{\chi_{Q0}} & = H^{\chi_{Q0}}\, \Big(
            {\cal C}\big[g_{1\sL}^g g_{1\sL}^g\big] - {\cal C}\big[w_{\scriptscriptstyle LL}\, h_{1\sL}^{\perp\,g} h_{1\sL}^{\perp\,g}\big] \Big)\, \langle 0\vert {\cal{O}}^{\chi_{Q0}}_1 (^3 P_0)\vert 0\rangle \, , \\
        F_{LL}^{\chi_{Q2}} & = -  H^{\chi_{Q2}}\, 
            {\cal C}\big[g_{1\sL}^g g_{1\sL}^g\big] \, \langle 0\vert {\cal{O}}^{\chi_{Q2}}_1 (^3 P_2)\vert 0\rangle \, ,
    \end{aligned}
\label{eq: FLL}
\end{equation}
with
\begin{equation}
    w_{\scriptscriptstyle LL} = 4\, w_{\scriptscriptstyle UU} = \frac{1}{M_p^4} \big[ 2\, (\bm p_{a\sT} \cdot \bm p_{b \sT})^2 - \bm p_{a\sT}^2 \bm p_{b\sT}^2 \big] = \frac{\bm p_{a\sT}^2 \bm p_{b\sT}^2}{M_p^4} \cos \big[ 2(\phi_a - \phi_b) \big] \, .
\end{equation}
We note that they turn out to be very similar to $F_{UU}$, with the role of the unpolarized gluon TMD taken by the helicity distribution $g_{1\sL}$.
Since the magnitude of the collinear $g_{1\sL}$ gluon distribution is much smaller than the unpolarized one, as suggested in Refs.~\cite{STAR:2006opb, STAR:2014wox, PHENIX:2014axc, PHENIX:2020trf, Zhou:2022wzm}, we expect that the same is valid also for its transverse momentum dependent counterpart. Hence, measurements of observables related to these convolutions are expected to be challenging.

Moving to the mixed double-polarized structure functions, we obtain
\begin{equation}
    \begin{aligned}
        F_{LT}^{\eta_Q, \cos \phi_{S_B}} & = H^{\eta_Q} \Big(
            - {\cal C}\big[w_{\scriptscriptstyle LT}^g\, g_{1 \sL}^g g_{1\sT}^{\perp\, g}\big] - {\cal C}\big[w_{\scriptscriptstyle LT}^{h}\, h_{1\sL}^{\perp\,g} h_1^{g}\big] - {\cal C}\big[w_{\scriptscriptstyle LT}^{h^\perp}\, h_{1\sL}^{\perp\,g} h_{1\sT}^{\perp\, g}\big] \Big)\, \langle 0\vert {\cal{O}}^{\eta_Q}_1 (^1 S_0)\vert 0\rangle \, , \\
        F_{LT}^{\chi_{Q0}, \cos \phi_{S_B}} & = H^{\chi_{Q0}} \Big(
            - {\cal C}\big[w_{\scriptscriptstyle LT}^g\, g_{1L}^g g_{1\sT}^{\perp\, g}\big] + {\cal C}\big[w_{\scriptscriptstyle LT}^{h}\, h_{1\sL}^{\perp\,g} h_1^{g}\big] + {\cal C}\big[w_{\scriptscriptstyle LT}^{h^\perp}\, h_{1\sL}^{\perp\, g} h_{1\sT}^{\perp\, g}\big] \Big)\, \langle 0\vert {\cal{O}}^{\chi_{Q0}}_1 (^3 P_0)\vert 0\rangle \, , \\
        F_{LT}^{\chi_{Q2}, \cos \phi_{S_B}} & = H^{\chi_{Q2}} \, 
             {\cal C}\big[w_{\scriptscriptstyle LT}^g\, g_{1 \sL}^g g_{1\sT}^{\perp\, g}\big] \, \langle 0\vert {\cal{O}}^{\chi_{Q2}}_1 (^3 P_2)\vert 0\rangle \, ,
    \end{aligned}
\label{eq: FLT}
\end{equation}
with
\begin{equation}
    \begin{aligned}
        w_{\scriptscriptstyle LT}^f & = w_{\scriptscriptstyle UT}^f
            = \frac{|\bm p_{b\sT}|}{M_p} \cos \phi_b \, , \\
        w_{\scriptscriptstyle LT}^h & = 2\, w_{\scriptscriptstyle UT}^h
            = \frac{\bm p_{a\sT}^2\, |\bm p_{b\sT}|}{2 M_p^3} \cos(\phi_b - 2 \phi_a) \, , \\
        w_{\scriptscriptstyle LT}^{h^\perp} & = 2\,w_{\scriptscriptstyle UT}^{h^\perp}
            = \frac{\bm p_{a\sT}^2\, |\bm p_{b\sT}|^3}{4 M_p^5} \cos(3 \phi_b - 2 \phi_a) \, ;
        \end{aligned}
\label{eq: wLT}
\end{equation}
and
\begin{equation}
    \begin{aligned}
        F_{TL}^{\eta_Q, \cos \phi_{S_A}} & = H^{\eta_Q} \Big(
            - {\cal C}\big[w_{\scriptscriptstyle TL}^g\, g_{1\sT}^{\perp\, g} g_{1\sL}^g\big] - {\cal C}\big[w_{\scriptscriptstyle TL}^{h}\, h_1^{g} h_{1\sL}^{g}\big] - {\cal C}\big[w_{\scriptscriptstyle TL}^{h^\perp}\, h_{1\sT}^{\perp\, g} h_{1\sL}^{g}\big] \Big)\, \langle 0\vert {\cal{O}}^{\eta_Q}_1 (^1 S_0)\vert 0\rangle \, , \\
        F_{TL}^{\chi_{Q0}, \cos \phi_{S_A}} & = H^{\chi_{Q0}} \Big(
            -{\cal C}\big[w_{\scriptscriptstyle TL}^g\, g_{1\sT}^{\perp\, g} g_{1\sL}^g\big] + {\cal C}\big[w_{\scriptscriptstyle TL}^{h}\, h_1^{g} h_{1\sL}^{g}\big] + {\cal C}\big[w_{\scriptscriptstyle TL}^{h^\perp}\, h_{1\sT}^{\perp\, g} h_{1\sL}^{g}\big] \Big)\, \langle 0\vert {\cal{O}}^{\chi_{Q0}}_1 (^3 P_0)\vert 0\rangle \, , \\
        F_{TL}^{\chi_{Q2}, \cos \phi_{S_A}} & = H^{\chi_{Q2}} \, 
             {\cal C}\big[w_{\scriptscriptstyle TL}^g\, g_{1\sT}^{\perp\, g} g_{1\sL}^g\big] \, \langle 0\vert {\cal{O}}^{\chi_{Q2}}_1 (^3 P_2)\vert 0\rangle \, ,
    \end{aligned}
\label{eq:FTL}
\end{equation}
where the weight functions in Eq.~\eqref{eq:FTL} can be obtained from Eq.~\eqref{eq: wLT} performing the substitutions $T\leftrightarrow L$ and $a \leftrightarrow b$.

The double-transverse polarization completes the picture. In this case, we find two separate structure functions corresponding to two different azimuthal modulations. Explicitly, these are:

\begin{equation}
    \begin{aligned}
        F_{TT}^{\eta_Q, \cos (\phi_{S_A} - \phi_{S_B})} & = H^{\eta_Q}\, \langle 0\vert {\cal{O}}^{\eta_Q}_1 (^1 S_0)\vert 0\rangle 
        \\ & \phantom{=}
        \times \Big( - {\cal C} \big[w_{\scriptscriptstyle TT}\, f_{1\sT}^{\perp\, g} f_{1\sT}^{\perp\, g}\big] + {\cal C}\big[w_{\scriptscriptstyle TT}\, g_{1\sT}^{\perp\, g} g_{1\sT}^{\perp\, g}\big]
        + {\cal C}\big[ w_{\scriptscriptstyle TT}^{h}\, h_{1}^{g} h_{1}^{g}\big] + {\cal C}\big[ w_{\scriptscriptstyle TT}^{h^\perp}\, h_{1\sT}^{\perp\, g} h_{1\sT}^{\perp\, g}\big]
            \Big)\, , \\
        F_{TT}^{\chi_{Q0}, \cos (\phi_{S_A} - \phi_{S_B})} & = H^{\chi_{Q0}}\, \langle 0\vert {\cal{O}}^{\chi_{Q0}}_1 (^3 P_0)\vert 0\rangle \\ & \phantom{=}
        \times \Big( - {\cal C}\big[w_{\scriptscriptstyle TT}^{f}\, f_{1\sT}^{\perp\, g} f_{1\sT}^{\perp\, g}\big] + {\cal C}\big[w_{\scriptscriptstyle TT}^{g}\, g_{1\sT}^{\perp\, g} g_{1\sT}^{\perp\, g}\big] - {\cal C}\big[ w_{\scriptscriptstyle TT}^{h}\, h_{1}^{g} h_{1}^{g}\big] - {\cal C}\big[ w_{\scriptscriptstyle TT}^{h^\perp}\, h_{1\sT}^{\perp\, g} h_{1\sT}^{\perp\, g}\big]\Big)  \, , \\
        F_{TT}^{\chi_{Q2}, \cos(\phi_{S_A} - \phi_{S_B})} & = H^{\chi_{Q2}} \Big( - {\cal C}\big[w_{\scriptscriptstyle TT}^{f}\, f_{1\sT}^{\perp\, g} f_{1\sT}^{\perp\, g}\big] - {\cal C}\big[w_{\scriptscriptstyle TT}^{g}\, g_{1\sT}^{\perp\, g} g_{1\sT}^{\perp\, g}\big]\Big)\, \langle 0\vert {\cal{O}}^{\chi_{Q2}}_1 (^3 P_2)\vert 0\rangle \, ,
    \end{aligned}
\label{eq: FTT 1}
\end{equation}
with
\begin{equation}
    \begin{aligned}
        w_{\scriptscriptstyle TT} & =  
        \frac{|\bm p_{a \sT}|\, |\bm p_{b \sT}|}{2 M_p^2} \cos (\phi_a - \phi_b)
        \, , \\
        w_{\scriptscriptstyle TT}^h & =  
        \frac{|\bm p_{a \sT}|\, |\bm p_{b \sT}|}{4 M_p^2} \cos (\phi_a - \phi_b)\, , \\
        w_{\scriptscriptstyle TT}^{h^\perp} & = 
        \frac{|\bm p_{a \sT}|^3 |\bm p_{b \sT}|^3}{16 M_p^6} \cos \big[ 3 (\phi_b - \phi_a) \big] \, ;
        \end{aligned}
\end{equation}
and
\begin{equation}
    \begin{aligned}
        F_{TT}^{\eta_Q, \cos(\phi_{S_A} + \phi_{S_B})} & = H^{\eta_Q}\, \langle 0\vert {\cal{O}}^{\eta_Q}_1 (^1 S_0)\vert 0\rangle  \\ & \phantom{=}
        \times \Big( {\cal C}\big[\overline w_{\scriptscriptstyle TT}\, f_{1\sT}^{\perp\, g} f_{1\sT}^{\perp\, g}\big] + {\cal C}\big[\overline w_{\scriptscriptstyle TT}\, g_{1\sT}^{\perp\, g} g_{1\sT}^{\perp\, g}\big] 
        + {\cal C}\big[ \overline w_{\scriptscriptstyle TT}^{hh^\perp}\, h_{1}^{g} h_{1\sT}^{\perp\, g}\big] + {\cal C}\big[ \overline w_{\scriptscriptstyle TT}^{h^\perp h}\, h_{1\sT}^{\perp\, g} h_{1}^{g}\big] \Big) \, , \\
        F_{TT}^{\chi_{Q0}, \cos (\phi_{S_A} + \phi_{S_B})} & = H^{\chi_{Q0}}\, \langle 0\vert {\cal{O}}^{\chi_{Q0}}_1 (^3 P_0)\vert 0\rangle \\ & \phantom{=}
        \times \Big( {\cal C}\big[\overline w_{\scriptscriptstyle TT}^{f}\, f_{1\sT}^{\perp\, g} f_{1\sT}^{\perp\, g}\big] + {\cal C}\big[\overline w_{\scriptscriptstyle TT}^{g}\, g_{1\sT}^{\perp\, g} g_{1\sT}^{\perp\, g}\big] - {\cal C}\big[\overline w_{\scriptscriptstyle TT}^{hh^\perp}\, h_{1}^{g} h_{1\sT}^{\perp\, g}\big] - {\cal C}\big[\overline w_{\scriptscriptstyle TT}^{h^\perp h}\, h_{1\sT}^{\perp\, g} h_{1}^{g}\big]\Big) \, , \\
        F_{TT}^{\chi_{Q2}, \cos(\phi_{S_A} + \phi_{S_B})} & = H^{\chi_{Q2}} \Big( {\cal C}\big[\overline w_{\scriptscriptstyle TT}^{f}\, f_{1\sT}^{\perp\, g} f_{1\sT}^{\perp\, g}\big] - {\cal C}\big[\overline w_{\scriptscriptstyle TT}^{g}\, g_{1\sT}^{\perp\, g} g_{1\sT}^{\perp\, g}\big] \Big)\, \langle 0\vert {\cal{O}}^{\chi_{Q2}}_1 (^3 P_2)\vert 0\rangle \, , \\
    \end{aligned}
\label{eq: FTT 2}
\end{equation}
with
\begin{equation}
    \begin{aligned}
        \overline w_{\scriptscriptstyle TT} & = 
        \frac{|\bm p_{a \sT}|\, |\bm p_{b \sT}|}{2 M_p^2} \cos (\phi_a + \phi_b) \, , \\
        \overline w_{\scriptscriptstyle TT}^{hh^\perp} & = 
        \frac{|\bm p_{a\sT}|\, |\bm p_{b\sT}|^3}{8 M_p^4} \cos (3 \phi_b - \phi_a) \, , \\
        \overline w_{\scriptscriptstyle TT}^{h^\perp h} & = 
        \frac{|\bm p_{a\sT}|^3\, |\bm p_{b\sT}|}{8 M_p^4} \cos (3 \phi_a - \phi_b) \, .
        \end{aligned}
\end{equation}
Unlike the previously discussed structure functions, $F^{{\cal Q}, \cos(\phi_{S_A} + \phi_{S_B})}_{TT}$ and $F^{{\cal Q}, \cos(\phi_{S_A} - \phi_{S_B})}_{TT}$ for $\chi_{Q2}$  production involve two different TMD convolutions.
However, since we expect to extract the gluon Sivers function from (transverse) SSAs, a measurement of the double-polarized structure functions for $J=2$ could be useful to isolate the contribution from $g_{1\sT}^{\perp\, g}$.
Interestingly, the two structure functions for the $J=0$ states provide complementary information: $F_{TT}^{\cos(\phi_{S_A} - \phi_{S_B})}$ is mainly sensitive to the magnitude of the distributions since it always involves the convolutions of the “square” of two gluon TMDs, whereas $F_{TT}^{\cos(\phi_{S_A} + \phi_{S_B})}$ could probe the relative sign of $h_1^g$ and $h_{1\sT}^{\perp\, g}$ because it contains the convolutions of these two TMDs.

We note that the convolutions in Eqs.~\eqref{eq: FUU}-\eqref{eq: FTT 2} are calculated in momentum space. In Appendix~\ref{app: convolution bT-space} we present instead, for a general spin-$1/2$ hadron, all the above convolutions in $b_\sT$-space (see Eqs.~\eqref{eq: convolution f1 f1}-\eqref{eq: convolution h1T h1}), where $\bm b_\sT$ is the Fourier conjugate of $\bm q_\sT$. Such expressions are particularly useful for future implementations of TMD evolution, which is multiplicative in $\bm b_\sT$-space. To this aim, a thorough calculation at the one-loop level of the so-called matching coefficients for all leading-twist gluon TMDs is underway.~\cite{Alvaro:2024ab}  

\section{Estimate of the upper bounds of the transverse single-spin asymmetries}
\label{sec: numerical results}

The angular modulations of the cross section for the process $pp\to {\cal Q} X$, presented in \cref{eq: diff cs full parameters},  can be singled out by taking average values of appropriate circular functions of $\phi_{S_A}$ and $\phi_{S_B}$, denoted as $W(\phi_{S_A}, \phi_{S_B})$,
\begin{align}
\langle W(\phi_{S_A}, \phi_{S_B})\rangle  = \frac{\int \d \phi_{S_A} \d \phi_{S_B} \, W(\phi_{S_A}, \phi_{S_B})\,   \d\sigma}{\int \d \phi_{S_A} \d \phi_{S_B} \,\d\sigma}\,, 
\end{align}
where we have used the notation $\d\sigma = \d\sigma/(\d y \, \d^2 q_\sT)$. In the following we focus on the specific configuration where only the proton with momentum $P_B$ is transversely polarized, whereas the proton with momentum $P_A$ is unpolarized, see Eq.~\eqref{eq:process}. Such processes could be in principle accessible at LHCSpin, the fixed target experiment planned at the LHC. In this case, we can define the azimuthal moments as
\begin{equation}
A_N^{{\cal Q}, \sin\phi_{S_B}} = 2\, \frac{\int \d\phi_{S_B} \sin\phi_{S_B} \left [\d\sigma (\phi_{S_B} ) - \d\sigma (\phi_{S_B} + \pi)\right ]}{\int \d\phi_{S_B} \left [\d\sigma (\phi_{S_B} ) +\d\sigma (\phi_{S_B} + \pi)\right ]} = \frac{F^{{\cal Q},\sin\phi_{S_B}}_{UT}}{F^{\cal Q}_{UU}} = 2 \, \langle \sin\phi_{S_B}\rangle\, ,
\label{eq: SSA A_N^SB}
\end{equation}
where we have assumed that the initial proton is fully polarized, namely $|\bm S_{BT}| = 1$, while $F^{\cal Q}_{UU}$ and $F^{{\cal Q},\sin\phi_{S_B}}_{UT}$ are given in Eqs.~\eqref{eq: FUU} and \eqref{eq:F_UT}, respectively.

\begin{figure}[t]
  \centering
  \includegraphics[width=0.4\linewidth, keepaspectratio]{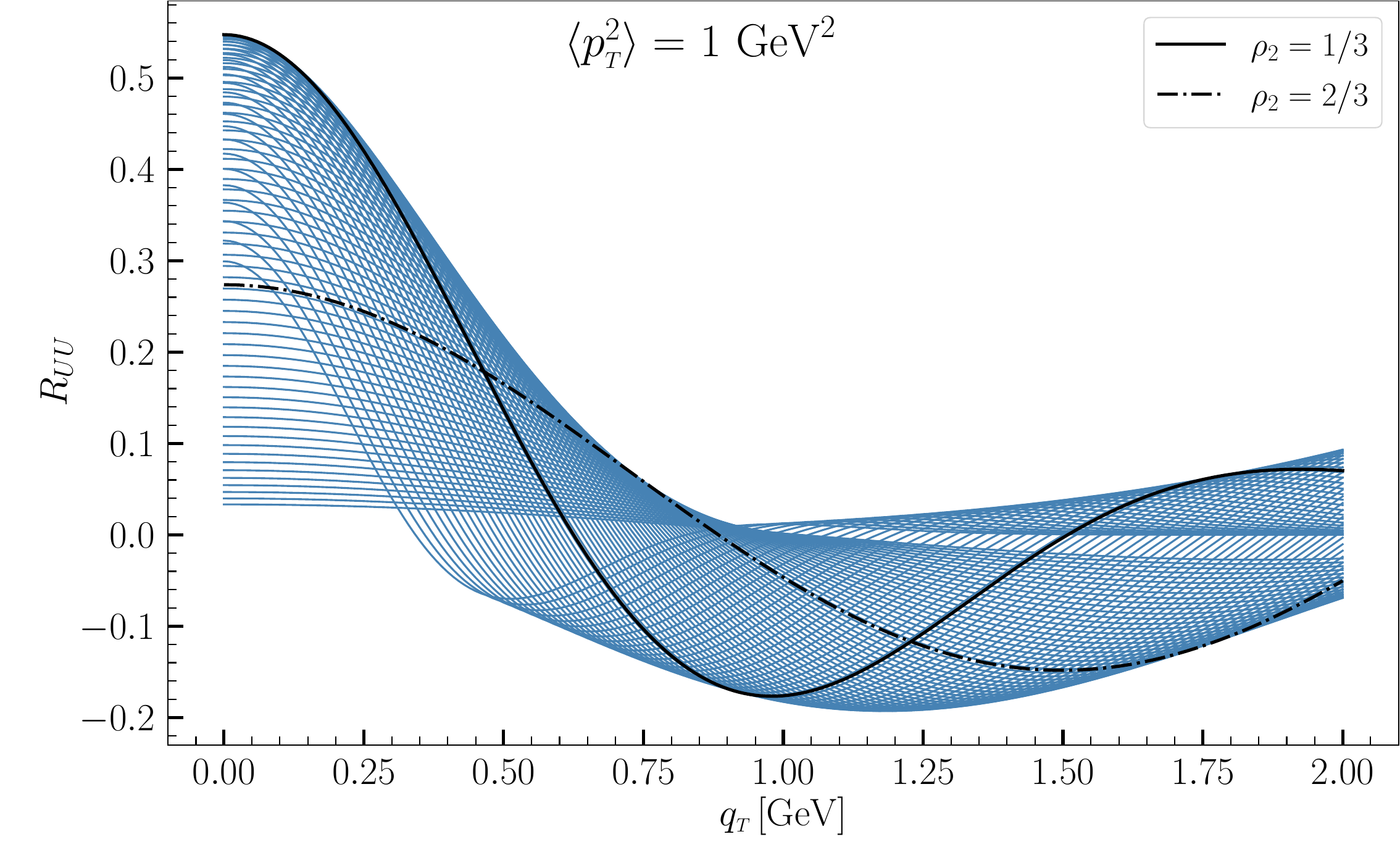}
  \caption{The ratio $R_{UU}$ as a function of $q_\sT$. The gluon TMDs $h_1^{\perp\, g}$ and $f_1^g$  are evaluated according to a Gaussian ansatz, with the average of the transverse momentum squared set at $1$~GeV$^2$. Different blue lines are obtained by varying the parameter $\rho_2$ in Eq.~\eqref{eq:RUU} from $0.1$ to $0.9$ with steps of $0.01$. We have highlighted the ratios obtained with two specific values of $\rho_2$, namely $\rho_2 = 1/3$ (solid black line) and $\rho_2 = 2/3$ (dashed-dotted line).}
  \label{fig: ratio UU}
\end{figure}

In order to provide an estimate of the upper bounds of the SSAs, we assume that the unpolarized gluon TMD has the following Gaussian form~\cite{Boer:2011kf,Lansberg:2017dzg},
\begin{equation}
f_1^g(x,\bm p_\sT^2) = \frac{f_1^g(x)}{\pi \langle  p_\sT^2 \rangle}\,
\exp\left[-\frac{\bm p_\sT^2}{\langle  p_\sT^2 \rangle}\right]\, ,
\label{eq: parameterization unpolar}
\end{equation}
with $f_1^g(x)$ being the collinear gluon distribution. The width $\langle p_\sT^2 \rangle$ 
could in principle depend on the energy scale, which is set by the quarkonium mass $M$. Furthermore, we take 
$\langle p_T^2\rangle$ to be independent of $x$. The effect of the other unknown TMDs will be maximal when
they saturate the following, model-independent, positivity bounds~\cite{Mulders:2000sh} 
\begin{equation}
\begin{aligned}
\big\vert f_{1\sT}^{\perp \,g}(x,\bm p_\sT^2) \big\vert,\ \big\vert h_{1}^g(x,\bm p_\sT^2)\big\vert & \le \frac{M_p}{\vert \pT \vert} f_1^g(x,\pT^2) \, , \\
\frac{1}{2}\, \vert h_{1}^{\perp \,g}(x,\bm p_\sT^2) \vert & \le \frac{M_p^2}{\pT^2}f_1^g(x,\pT^2) \, , \\
\frac{1}{2}\, \vert h_{1\sT}^{\perp\, g}(x,\bm p_\sT^2) \vert & \le \frac{M_p^3}{\vert \pT \vert^3} f_1^g(x,\pT^2) \, .
\end{aligned}
\end{equation}
These bounds are always fulfilled, although not everywhere saturated, if we take, along the lines of Refs.~\cite{Boer:2011kf,Boer:2012bt},
\begin{equation}
\begin{aligned}
f_{1\sT}^{\perp\,g}(x,\bm p_\sT^2) & = 
    {\cal N}_0(x)\, \frac{f_1^g(x)}{\pi \langle p_\sT^2 \rangle^{3/2}}\, M_p \, \sqrt{\frac{2 (1 - \rho_0)}{\rho_0}}
    \exp\left[\frac{1}{2} - \frac{1}{\rho_0}\,\frac{\bm p_\sT^2}{\langle p_\sT^2 \rangle}\right] \, , \\
h_{1}^{g}(x,\bm p_\sT^2) & = 
    {\cal N}_1(x)\, \frac{f_1^g(x)}{\pi \langle p_\sT^2 \rangle^{3/2}}\, M_p \, \sqrt{\frac{2 (1 - \rho_1)}{\rho_1}}
    \exp\left[\frac{1}{2} - \frac{1}{\rho_1}\,\frac{\bm p_\sT^2}{\langle p_\sT^2 \rangle}\right] \, , \\
h_{1}^{\perp\,g}(x,\bm p_\sT^2) & = 
    2\, {\cal N}_2(x)\, \frac{f_1^g(x)}{\pi \langle p_\sT^2 \rangle^{2}}\, M_p^2 \, {\frac{(1 - \rho_2)}{\rho_2}} 
    \exp\left[1 - \frac{1}{\rho_2}\, \frac{\bm p_\sT^2}{\langle p_\sT^2 \rangle}\right] \, , \\
h_{1\sT}^{\perp\,g}(x,\bm p_\sT^2) & = 
    2\, {\cal N}_3(x)\, \frac{f_1^g(x)}{\pi \langle p_\sT^2 \rangle^{5/2}}\, M_p^3 \, \left [\frac{2 (1 - \rho_3)}{3 \rho_3} \right ]^{3/2} \,
    \exp\left[\frac{3}{2} - \frac{1}{\rho_3}\,\frac{\bm p_\sT^2}{\langle p_\sT^2 \rangle}\right] \, , \\
\label{eq: parameterization asymmetries}
\end{aligned}
\end{equation}
where the free parameters $\rho_i$ are such that $0 < \rho_i < 1$ and
\begin{equation}
    {\cal N}_i(x) = N_i\ x^{\alpha_i} (1 - x)^{\beta_i} \frac{(\alpha_i + \beta_i)^{\alpha_i + \beta_i}}{\alpha_i^{\alpha_i} \beta_i^{\beta_i}},
\end{equation}
with $\vert N_i \vert \leq 1$.
The functions ${\cal N}_i(x)$ account for a different $x$-dependence of the gluon TMDs in \cref{eq: parameterization asymmetries} with respect to the unpolarized one.
Since in this section we aim at providing only the upper bounds of $A_N^{\sin\phi_{S_B}}$, we take ${\cal N}_i(x) = 1$ and let the $x$-dependence saturate the positivity bounds. Moreover, we note that all TMDs are taken to be positive.

Hence, the ratios of TMD convolutions entering the SSAs are given by
\begin{align}
    R_{UU} & = 
    {\cal C}\left[w^h_{\scriptscriptstyle UU}\, h_{1}^{\perp\, g}\, h_{1}^{\perp\, g} \right] / {\cal C} \left[ f_1^g\,f_1^g \right] \nonumber \\ & =
        \frac{1}{16\, \langle p_\sT^2 \rangle^2 } \frac{(1 - \rho_2)^2}{\rho_2} \Big(\bm q_\sT^4 - 8\, \rho_2\,\langle p_\sT^2 \rangle\, \bm q_\sT^2 + 8\,\rho_2^2\, \langle p_\sT^2 \rangle^2 \Big)\, \exp{\left[2- \frac{1 - \rho_2}{\rho_2} \frac{\bm q_\sT^2}{2\,\langle p_\sT^2 \rangle}\right]}\, , \label{eq:RUU}\\
    R_{UT}^f & = 
    {\cal C} \left[w^f_{\scriptscriptstyle UT}\, f_1^g\, f_{1\sT}^{\perp\,g} \right]/ {\cal C} \left[ f_1^g\,f_1^g \right] \nonumber \\ & =
        \frac{2}{\langle p_\sT^2 \rangle^{1/2}}\sqrt{\frac{2(1 - \rho_0)}{\rho_0}} \Big(\frac{\rho_0}{1+\rho_0}\Big)^2 \vert \bm q_\sT \vert \exp{\left[\frac{1}{2}-\frac{1-\rho_0}{1 + \rho_0}\frac{\bm q_\sT^2}{2\,\langle p_\sT^2 \rangle}\right]}\, , \\
    R_{UT}^h & = 
    {\cal C} \left[w^h_{\scriptscriptstyle UT}\, h_1^{\perp\, g}\, h_{1}^{g} \right]/ {\cal C} \left[ f_1^g\,f_1^g \right] \nonumber \\ & =
        \frac{1}{\langle p_\sT^2 \rangle^{3/2}}\sqrt{\frac{2(1 - \rho_1)}{\rho_1}} (1 - \rho_2)\, \frac{{\rho_1}^2\, \rho_2^2}{\big(\rho_1 + \rho_2 \big)^4}\, \vert \bm q_\sT \vert\, \Big( \bm q_\sT^2 - 2 (\rho_1 + \rho_2) \langle p_\sT^2 \rangle \Big) \exp{\left[\frac{3}{2}-\frac{2-\rho_1-\rho_2}{\rho_1+\rho_2}\frac{\bm q_\sT^2}{2\,\langle p_\sT^2 \rangle}\right]}\, , \\
    R_{UT}^{h^\perp} & = 
    {\cal C} \left[w^{h^\perp}_{\scriptscriptstyle UT}\, h_1^{\perp\, g}\, h_{1\sT}^{\perp\, g} \right]/ {\cal C} \left[ f_1^g\,f_1^g \right] 
    \nonumber \\ & =
        \frac{1}{\langle p_\sT^2 \rangle^{5/2}}\left[\frac{2(1 - \rho_3)}{3 \rho_3}\right]^{3/2} (1 - \rho_2)\, \frac{\rho_2^2\, \rho_3^4}{\big(\rho_2 + \rho_3 \big)^6}\,
        \nonumber\\ & \phantom{=} \times
        \vert \bm q_\sT \vert\, \Big( \bm q_\sT^4 - 6\, (\rho_2 + \rho_3)\, \langle p_\sT^2 \rangle\, \bm q_\sT^2 + 6\, (\rho_2 + \rho_3)^2\, \langle p_\sT^2 \rangle^2 \Big) \exp{\left[\frac{5}{2} - \frac{2-\rho_2 - \rho_3}{\rho_2 + \rho_3}\frac{\bm q_\sT^2}{2\,\langle p_\sT^2 \rangle}\right]} \, ,
\end{align}
with each of them varying between zero and one.
Note that, while $R_{UU} \neq 0$ at $q_\sT = 0$, $R_{UT} = 0$ independently of the convolution considered. The azimuthal moments in Eq.~\eqref{eq: SSA A_N^SB} can therefore be rewritten as
\begin{equation}
\begin{aligned}
A_N^{\eta_Q, \sin\phi_{S_B}} & = \frac{-R_{UT}^f + R_{UT}^h - R_{UT}^{h^\perp}}{1-R_{UU}}\,,  \\
A_N^{\chi_{Q0}, \sin\phi_{S_B}} & = \frac{-R_{UT}^f - R_{UT}^h + R_{UT}^{h^\perp}}{1 + R_{UU}}\,,\\
A_N^{\chi_{Q 2}, \sin\phi_{S_B}} & = -R_{UT}^f \,.
\end{aligned}    
\label{eq:ANs}
\end{equation}

\begin{figure}[t]
  \centering    
  \hspace*{-0.4cm}
  \subfloat[\label{subfig:a}]{\includegraphics[width=.31\linewidth, keepaspectratio]{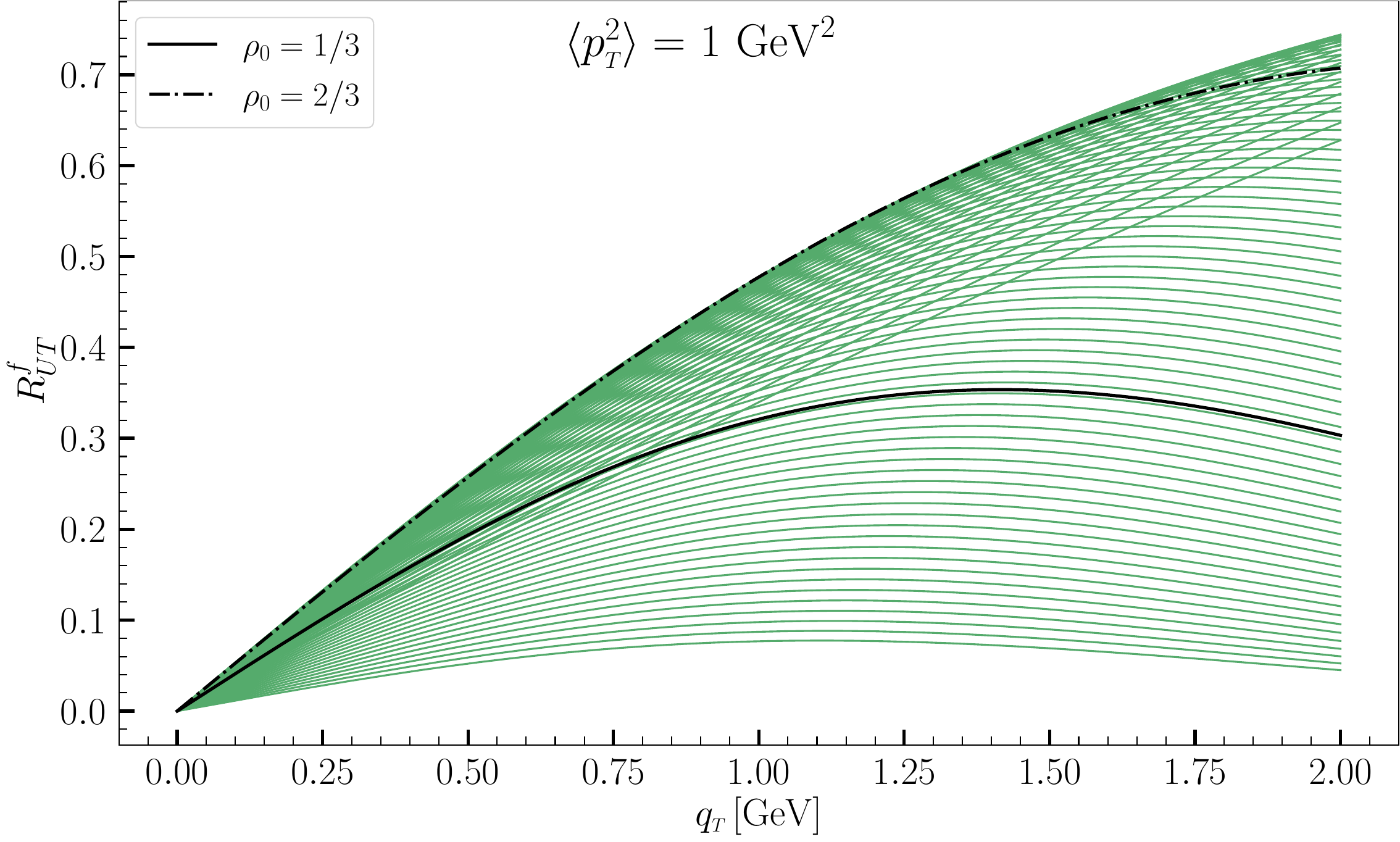}}\hspace*{0.5cm}
  \subfloat[\label{subfig:b}]{\includegraphics[width=.31\linewidth, keepaspectratio]{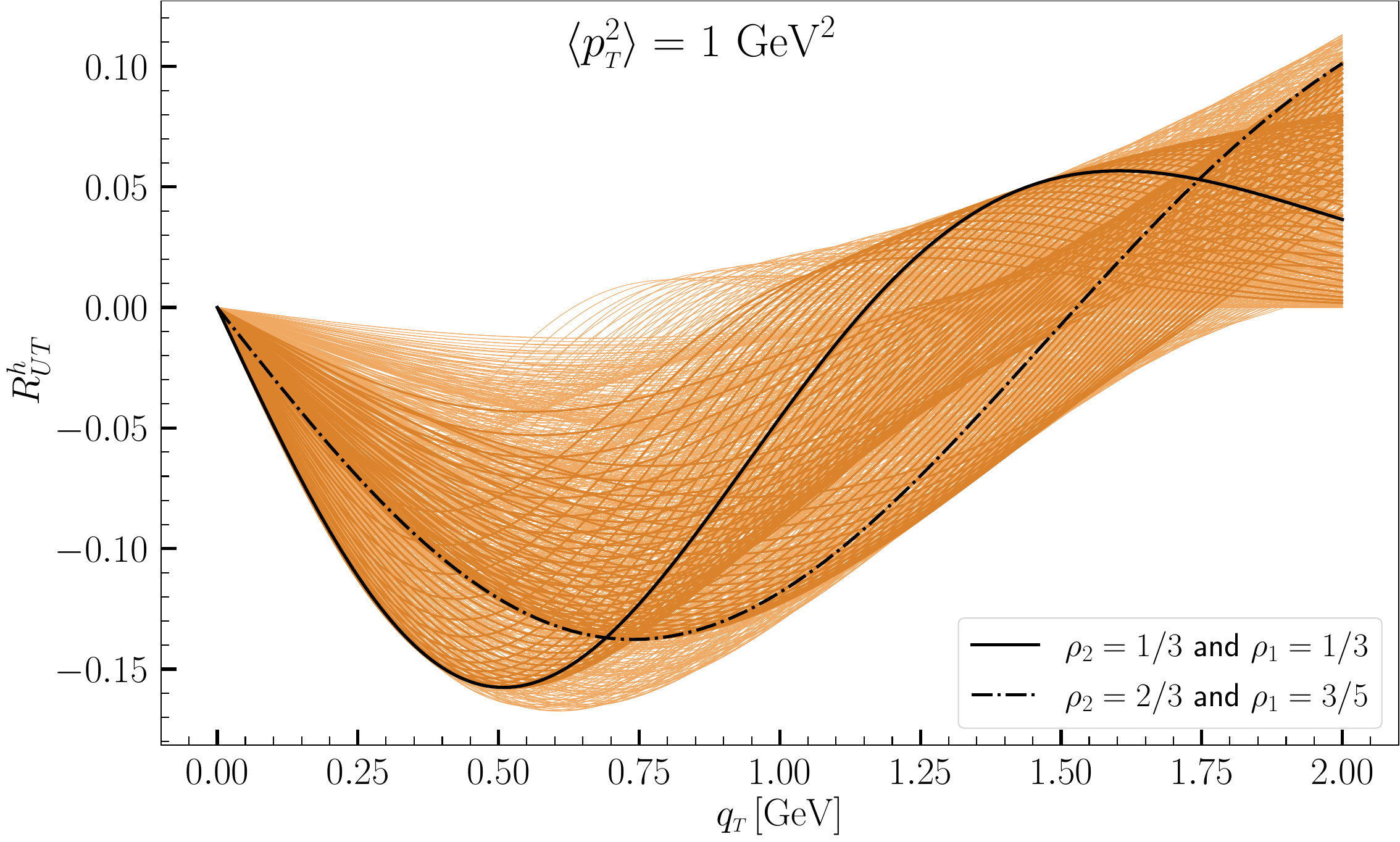}}\hspace*{0.5cm}
  \subfloat[\label{subfig:c}]{\includegraphics[width=.31\linewidth, keepaspectratio]{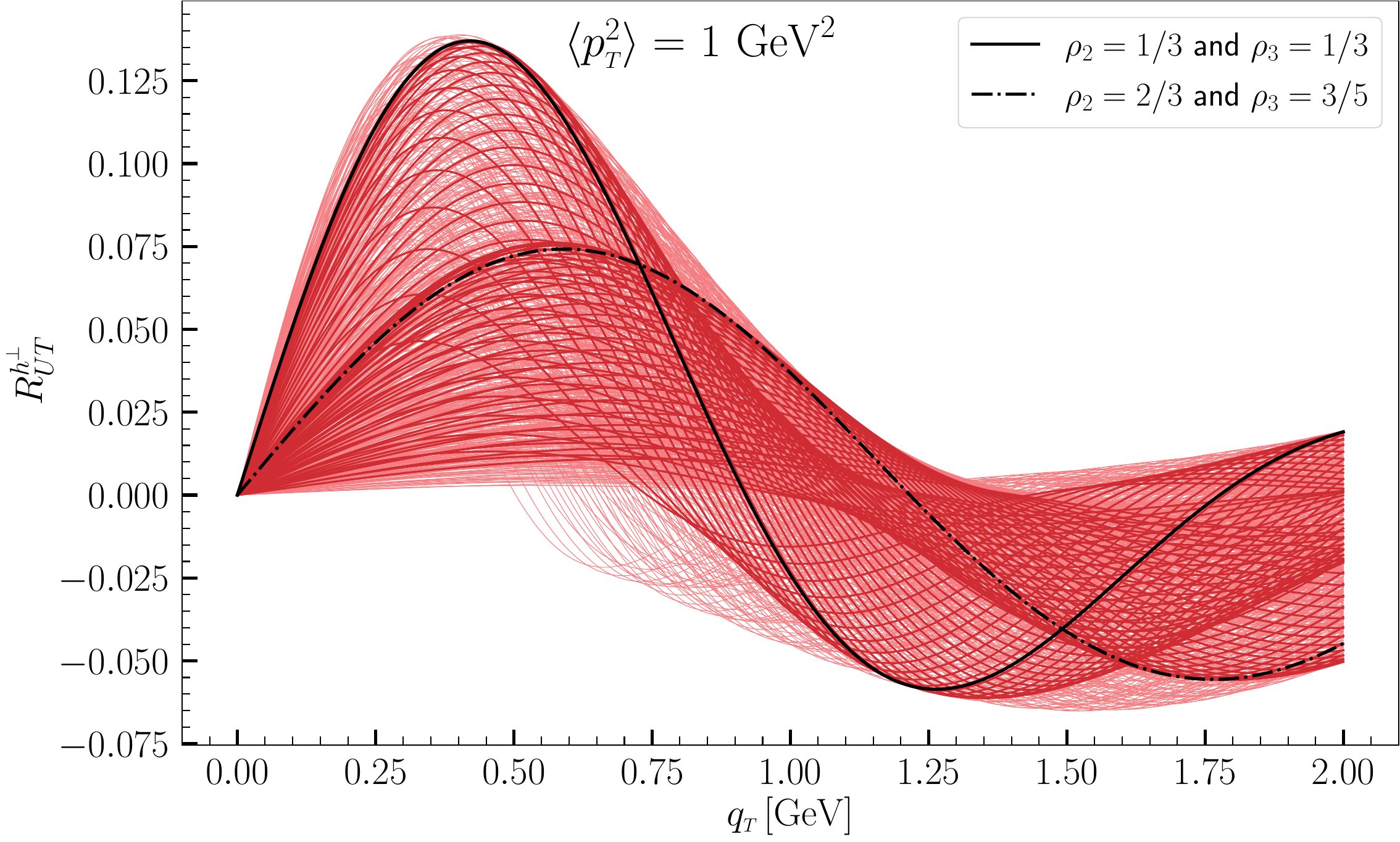}}
  \caption{The three ratios contributing to the numerators of the transverse SSAs as a function of $q_\sT$. The TMDs are evaluated according to a Gaussian ansatz, with the average of the transverse momentum squared set at $1$~GeV$^2$. Different curves are drawn by varying each parameter in its space. In particular: in a) the green lines correspond to $0.1 < \rho_0 < 0.9$ with steps of $0.01$; in b) the thicker orange lines are obtained by taking $\rho_2 = 1/3$ and $\rho_2 = 2/3$ and varying $0.1 < \rho_1 < 0.9$ with steps of $0.01$, while for the thinner ones we have varied both $\rho_2$ and $\rho_1$ independently with larger steps (for visualization reasons); in c) thicker and thinner red lines are evaluated as in b), but with $\rho_3$ replacing $\rho_1$. 
  Moreover, in each of the three figures we have highlighted two curves in black that correspond to the ratios obtained with two specific choices of the parameters (see the legend boxes in each figure).}
  \label{fig: ratio UT}
\end{figure}

In the following numerical study we focus on charmonium production. From existing phenomenological analyses, the value ${\langle p_\sT^2 \rangle = 1\ {\rm GeV}^2}$ turns out to be a reasonable choice for the Gaussian width of the unpolarized gluon TMD at the scale $\mu^2 = 4 M_c^2$~\cite{DAlesio:2017rzj, DAlesio:2019gnu, DAlesio:2020eqo}. Because of TMD evolution of the gluon densities, larger values of ${\langle p_\sT^2 \rangle}$ are expected for bottomonium production. Moreover, we show our prediction for $q_\sT \leq 2~{\rm GeV}$, to guarantee that our analysis is restricted in the kinematic region where TMD factorization is expected to be applicable.

In Fig.~\ref{fig: ratio UU} the ratio $R_{UU}$, contributing to the transverse momentum spectrum of the unpolarized cross section and the denominators of the SSAs for (pseudo)scalar quarkonia, is shown as a function of $q_\sT$ and for different values of the parameter $\rho_2$, in the range $0.1 < \rho_2 < 0.9$. Similarly, $R_{UT}^f$, $R_{UT}^h$ and $R_{UT}^{h^\perp}$, appearing in the numerators of the SSAs for  $\eta_Q$ and $\chi_{Q0}$ are presented in Fig.~\ref{fig: ratio UT} for several values of $\rho_0,~\rho_1,~\rho_2,~\rho_3$, as described in the caption.
These quantities measure the relative magnitude of the linearly polarized distributions and the Sivers function to the unpolarized gluon TMD. We note that our predictions are rather stable with respect to the choice of $\langle p_\sT^2\rangle$: the curves are shrunk to lower values of $q_\sT$ as $\langle p_\sT^2\rangle$ decreases, while they broaden as $\langle p_\sT^2\rangle$ increases, with the size of the ratios staying practically unchanged.

Our main results on the upper bounds of the azimuthal moments in Eq.~\eqref{eq:ANs} for the $J=0$ and $J=2$ charmonium states are shown in \cref{fig: AN} as a function of $q_\sT$. The parameters $\rho_i$ have been chosen in order to maximize the asymmetries in the TMD region. The red full lines indicate the SSAs for $\chi_{Q2}$ production, which are entirely driven by the gluon Sivers function.  By comparing the SSAs for $\eta_Q$ (green dashed lines) and $\chi_{Q0}$ states (blue dash-dotted lines) with those for $\chi_{Q2}$ it would be possible, in principle, to assess the relevance of the combined effects of the linearly polarized gluon TMDs. The additional modulations in the asymmetries for (pseudo)scalar mesons as compared to the ones for spin-2 states are due to the presence of $h_1^g$ and $h_{1T}^{\perp\, g}$ in the numerators, but also of $h_1^{\perp\,g}$ in the denominators. For this reason, we also present the former asymmetries with $h_1^g = h_{1\sT}^{\perp\, g} = 0$, corresponding to the pink dashed ($\eta_Q$) and gray dash-dotted ($\chi_{Q2}$) lines in \cref{fig: AN}. A comparison between either the pink and green dashed lines for $\eta_Q$ or the gray and blue dash-dotted lines for $\chi_{Q0}$ displays that the effect from $h_1^g$ and  $h_{1\sT}^{\perp\, g}$ may still be experimentally accessible, in case the data will show modulations exceeding the ones expected from $h_1^{\perp\, g}$. Moreover, according to our Gaussian model, the maximized impact of $h_1^g$ and $h_{1\sT}^{\perp\, g}$ is significantly large in the kinematic region $q_\sT \leq 1$ GeV, which is therefore expected to play an important role in accessing these completely unknown gluon TMD distributions.

\begin{figure}[t]
  \centering
  \includegraphics[width=0.75\linewidth, keepaspectratio]{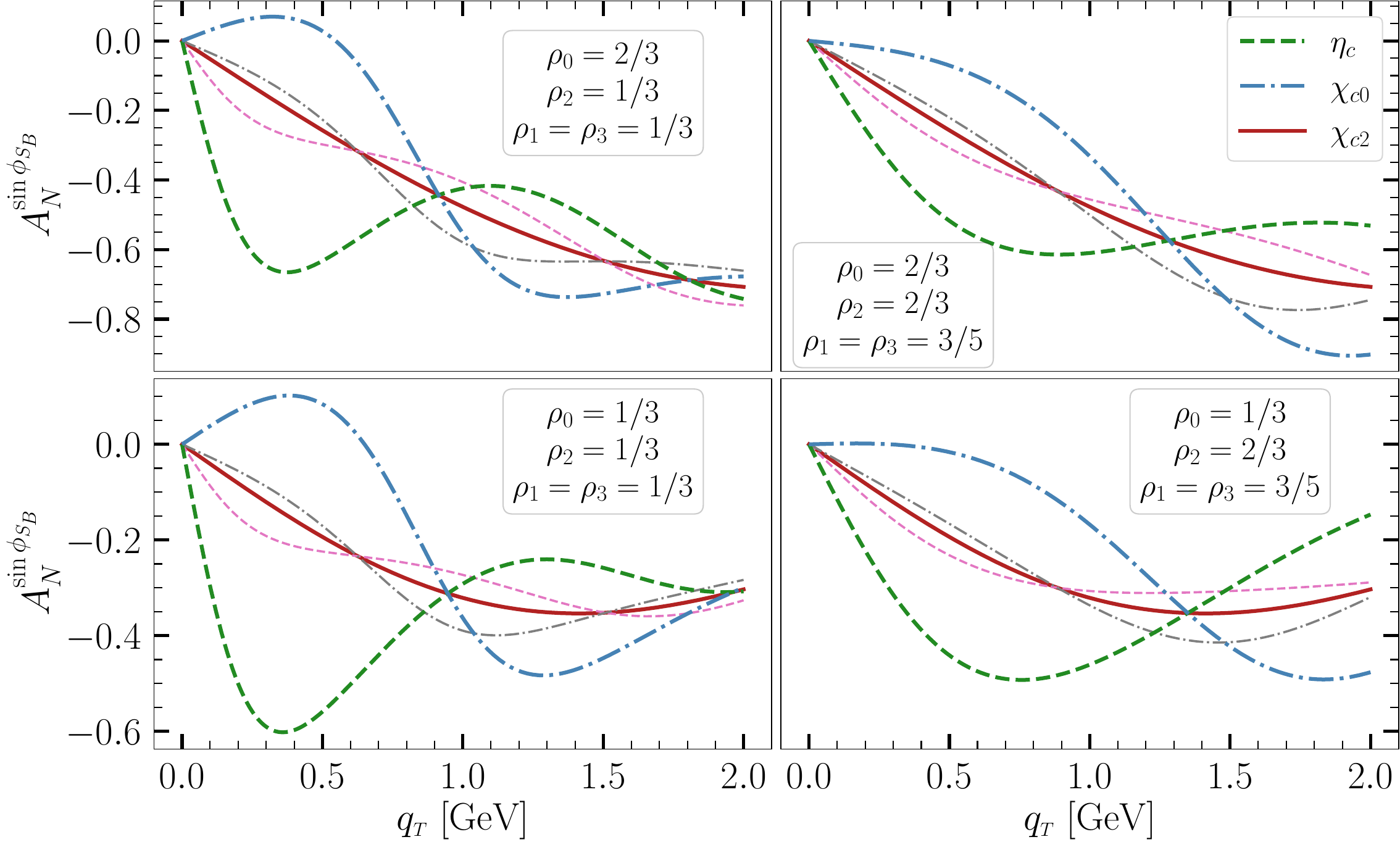}
  \caption{SSAs for different $C$-even quarkonia as a function of ${q_\sT}$.
  The gluon TMDs are evaluated according to Eqs.~\eqref{eq: parameterization unpolar} and~\eqref{eq: parameterization asymmetries} for different values of their parameters $\rho_i$ (check the text boxes in each panel for more details). The pink dashed and gray dash-dotted lines correspond to the SSAs of $\eta_Q$ and $\chi_{Q0}$, respectively, with the TMDs $h_1^g$ and $h_{1\sT}^{\perp\, g}$ set to $0$.}
  \label{fig: AN}
\end{figure}

\section{Summary and conclusions}
\label{sec:conclusions}

In this paper we have investigated the production of $C$-even quarkonium states in (un)polarized proton-proton collisions within the framework of TMD factorization. Supported by NRQCD arguments, we have adopted the color-singlet model to describe the quarkonium formation mechanism. We have derived the analytical expressions for the azimuthal modulations of the cross sections, arising from the convolutions of different leading-twist gluon TMDs. We therefore suggest that a phenomenological investigation of these quantities would provide direct access to the WW-type gluon distributions, in strong analogy with the studies of the Drell-Yan processes for the extraction of quark TMDs~\cite{Arnold:2008kf}. 

A striking, model-independent feature of our results is that all the linearly polarized gluon TMDs inside unpolarized, longitudinally and transversely polarized protons, namely $h_1^{\perp\, g}, h_{1\sL}^{\perp\, g}, h_1^g$ and $h_{1\sT}^{\perp\,g}$, contribute to the production of parity-odd $\eta_{Q}$ states 
with opposite signs with respect to the parity-even $\chi_{c, b 0}$ states. On the other hand, their effects on higher angular momentum quarkonia like $\chi_{Q 2}$ are strongly suppressed. As a consequence, as already pointed out in Ref.~\cite{Boer:2012bt}, by only looking at unpolarized scattering, the $\chi_{Q 2}$ cross section could be used to probe the distribution of unpolarized gluons $f_1^g$, while a combined study of $\eta_{Q}$ and $\chi_{Q0}$ would shed light on $h_1^{\perp\,g}$.  

Moreover, by employing simple Gaussian parameterizations for the gluon TMDs, which fulfill without saturating everywhere the well-known positivity bounds, we have estimated the maximal values of the transverse single-spin asymmetries, showing that they could be measured in principle at LHCSpin, the fixed target experiment planned at the LHC. Such observables for $\chi_{Q 2}$ are driven by the gluon Sivers function 
 $f_{1T}^{\perp\, g}$. Once this is known, SSAs for $\eta_Q$ and $\chi_{Q0}$ can be used to determine $h_1^g$ and $h_{1\sT}^{\perp\,g}$. Measurements of other observables, such as transverse and longitudinal double-spin asymmetries would be needed to have a full knowledge of the gluon distributions of the proton.
 
\section*{Acknowledgements}
We would like to thank Tomaso Pitzalis for his fundamental contribution during the early stages of this project. C.P.\ is indebted to Miguel Echevarria, Jean-Philippe Lansberg, Marc Schlegel and Andrea Signori for useful discussions. This work is supported by the European Union ``Next Generation EU'' program through the Italian PRIN 2022 grant n.~20225ZHA7W and by Fondazione di Sardegna
under the project ``Matter-antimatter asymmetry and polarisation in strange hadrons at LHCb'', project number F73C22001150007 (University of Cagliari).

\appendix
\section{TMD convolutions in Fourier \texorpdfstring{$b_\sT$}{\it b{\textunderscore}T}-space}
\label{app: convolution bT-space}
In this appendix we present the gluon correlator and the convolutions of TMDs in $b_\sT$-space, with $\bm b_\sT$ being the Fourier conjugate of the transverse momentum $\bm p_\sT$.  The Fourier transform of the correlator is defined as
\begin{align}
\tilde \Gamma_g^{\mu\nu} (x, \bm b_\sT) \equiv \int \d^2 \bm p_\sT \, e^{i\bm b_\sT \cdot \bm p_\sT}\, \Gamma_g^{\mu\nu} (x, \bm p_\sT)\,,
\end{align}
which leads to~\cite{Boer:2011xd, Boer:2016xqr}
\begin{equation}
\begin{aligned}
    2 \,x \,\tilde \Gamma^{\mu\nu}_{g\, U} (x, \bm b_\sT) & = - g_\sT^{\mu\nu} \tilde f_{1}^g(x,\bm b_\sT^2) - \frac{1}{2}\, M_h^2 \left ( b_\sT^\mu b_\sT^\nu + g_\sT^{\mu\nu} \frac{\bm b_\sT^2}{2} \right) \tilde h_1^{\perp\,g\,(2)}(x, \bm b_\sT^2) \, , \\
    2 \,x \,\tilde \Gamma^{\mu\nu}_{g\, L} (x, \bm b_\sT) & = S_\sL \left \{ i\, \epsilon_\sT^{\mu\nu}\tilde g_{1\sL} (x, \bm b_\sT^2) - 
    \frac{M_h^2}{2}\, \epsilon_\sT^{b_\sT \{\mu} b_\sT^{\nu\}}\, \tilde h_{1\sL}^{g\,(2)}(x, \bm b_\sT^2) \right \} \, ,  \\
    2 \,x\, \tilde \Gamma^{\mu\nu}_{g\, T} (x, \bm b_\sT) & = i \,M_h \, g_\sT^{\mu\nu} \epsilon_\sT^{b_\sT S_\sT} \, \tilde{f}_{1\sT}^{\perp\, g\, (1)} (x, \bm b_\sT^2) - M_h \, (S_\sT \cdot b_\sT)\, \tilde{g}_{1\sT}^{\perp\,g\,(1)}(x, \bm b_\sT^2) \\
    & \phantom{=} - i\,M_h\, \frac{b_{\sT\rho}\,\epsilon_\sT^{\rho\{\mu} S_\sT^{\nu\}} + S_{\sT \rho} \epsilon_\sT^{\rho \{\mu}b_\sT^{\nu\}}}{4}\,\tilde h_{1}^{g\,(1)}(x,\bm b_\sT^2) \\
    & \phantom{=} - i \,M_h^3\, \frac{4\,(b_\sT\cdot S_\sT)\,\epsilon_\sT^{b_\sT \{ \mu}b_\sT^{\nu \}} + \bm b_\sT^2 \left [ \epsilon_\sT^{b_\sT \{ \mu}S_\sT^{\nu \}} + 
    \epsilon_\sT^{S_\sT\{ \mu } b_\sT^{\nu \} }\right]}{48}\, \tilde h_{1 \sT}^{\perp\,g\,(3)}(x, \bm b_\sT^2)\, ,
\end{aligned}
\end{equation}
where we recall that $M_h$ is the mass of a general spin-$1/2$ hadron ($M_h = M_p$ for protons). Furthermore, we have introduced the Fourier transform of the generic TMD $f$
\begin{equation}
\tilde{f}(x, \bm b_\sT^2) \equiv \int \d^2 \bm p_\sT\, e^{i\bm b_\sT \cdot \bm p_\sT}f(x, \bm p_\sT^2)
= 2 \pi \int_0^\infty \d |\bm p_\sT| \,|\bm p_\sT|\, J_0(|\bm b_\sT| |\bm p_\sT|)\,f(x, \bm p_\sT^2) \,,
\label{eq: PDF FT}
\end{equation}
and its derivatives with respect to $b_\sT^2$
\begin{equation}
\tilde{f}^{(n)} (x, \bm b_\sT^2) \equiv  n! \left( -\frac{2}{M_h^2}\, \frac{\partial}{\partial \bm b_\sT^2} \right )^n \tilde f(x, \bm b_\sT^2) = \frac{2\pi n!}{(M_h^2)^n}
\int_0^{\infty} \d |\bm p_\sT|\, |\bm p_\sT| \left (\frac{ |\bm p_\sT|}{|\bm b_\sT|} \right )^n J_n(|\bm b_\sT| |\bm p_\sT|) 
f(x, \bm p_\sT^2)\, .
\label{eq: PDF FT derivatives}
\end{equation}
Moreover, $J_n$ in the above equations is the Bessel function of the first kind of order $n$, defined as
\begin{align}
J_n(z) = \frac{1}{2\pi i^n}\int_0^{2\pi}\d \varphi \,e^{i n \varphi}e^{i z \cos\varphi}\,.
\end{align}

By means of Eqs.~\eqref{eq: PDF FT} and \eqref{eq: PDF FT derivatives}, we find the following convolutions in $b_\sT$-space
\begin{align}
{\cal C}[f_1^g \,f_1^g] 
    & = \frac{1}{2\pi} \int_{0}^{\infty} \d |\bm b_\sT|\, |\bm b_\sT| \, J_0(|\bm b_\sT| |\bm q_\sT|) \, \tilde{f}_1^g(x_a, \bm b_\sT^2) \, \tilde{f}_1^g(x_b, \bm b_\sT^2) \, , 
\label{eq: convolution f1 f1}\\
{\cal C}[w_{\scriptscriptstyle UU} \, h_1^{\perp\, g}\, h_1^{\perp\,g}] 
    & = \frac{M_h^4}{32\pi} \int_0^\infty \d |\bm b_\sT| \, |\bm b_\sT|^5\, J_0 (|\bm b_\sT| |\bm q_\sT|)\,  \tilde{h}_1^{\perp\, g\, (2)}(x_a, \bm b_\sT^2)\, \tilde{h}_1^{\perp\, g\, (2)}(x_b, \bm b_\sT^2) \, , \\
{\cal C}\left[ w^f_{\scriptscriptstyle UT}\, f_1^g\, f_{1 \sT}^{\perp\, g} \right] 
    & = \frac{M_h}{2\pi} \int_{0}^{\infty} \d |\bm b_\sT| \, \bm b_\sT^2 \,  J_1(|\bm b_\sT| |\bm q_\sT|)\, \tilde{f}_1^g(x_a, \bm b_\sT^2) \, \tilde{f}_{1 \sT}^{\perp\, g\, (1)}(x_b, \bm b_\sT^2) \, , \\
{\cal C}\left[ w^h_{\scriptscriptstyle UT}\, h_1^{\perp\, g}\, h_1^g \right] 
    & = -\frac{M_h^3}{16 \pi} \int_0^\infty \d |\bm b_\sT| \, \bm b_\sT^4\, J_1(|\bm b_\sT| |\bm q_\sT|)\, \tilde{h}_1^{\perp\,g\, (2)}(x_a, \bm b_\sT^2)\, \tilde{h}_1^{g\, (1)}(x_b, \bm b_\sT^2)\, , \\
{\cal C}\left [w^{h^\perp}_{\scriptscriptstyle UT}\, h_1^{\perp\, g}\, h_{1 \sT}^{\perp\, g} \right] 
    & = \frac{M_h^5}{192\pi} 
 \int_0^\infty \d |\bm b_\sT|\, \bm b_\sT^6\, J_1(|\bm b_\sT| |\bm q_\sT|)\, \tilde{h}_1^{\perp\, g\, (2)}(x_a, \bm b_\sT^2) \, \tilde{h}_{1\sT}^{\perp\, g\, (3)}(x_b, \bm b_\sT^2) \, , \\
{\cal C}[g_{1 \sL}^g \, g_{1 \sL}^g] 
    & = \frac{1}{2\pi} \int_{0}^{\infty} \d |\bm b_\sT|\, |\bm b_\sT| \, J_0(|\bm b_\sT| |\bm q_\sT|) \, \tilde{g}_{1\sL}^g(x_a, \bm b_\sT^2) \, \tilde{g}_{1\sL}^g(x_b, \bm b_\sT^2) \, , \\
{\cal C}[w_{\scriptscriptstyle LL} \, h_{1\sL}^{g}\, h_{1\sL}^{g}] 
    & = \frac{M_h^4}{8\pi} \int_0^\infty \d |\bm b_\sT| \, |\bm b_\sT^5|\, J_0 (|\bm b_{\sT}| |\bm q_\sT|)\,  \tilde{h}_{1\sL}^{g\, (2)}(x_a, \bm b_\sT^2)\, \tilde{h}_{1\sL}^{g\, (2)}(x_b, \bm b_\sT^2) \, , \\
{\cal C}\left[ w^g_{\scriptscriptstyle LT}\, g_{1 \sL}^g\, g_{1 \sT}^{\perp\, g} \right] 
    & = \frac{M_h}{2\pi} \int_{0}^{\infty} \d |\bm b_\sT| \, \bm b_\sT^2 \,  J_1(|\bm b_\sT| |\bm q_\sT|) \, \tilde{g}_{1\sL}^g(x_a, \bm b_\sT^2) \, \tilde{g}_{1 \sT}^{\perp\, g\, (1)}(x_b, \bm b_\sT^2) \, , \\
{\cal C}\left[ w^h_{\scriptscriptstyle LT}\, h_{1\sL}^{g}\, h_1^g \right] 
    & = -\frac{M_h^3}{8 \pi} \int_0^\infty \d |\bm b_\sT| \,\bm b_\sT^4\, J_1(|\bm b_\sT| |\bm q_\sT|) \, \tilde{h}_{1\sL}^{g\, (2)}(x_a, \bm b_\sT^2)\, \tilde{h}_1^{g\, (1)}(x_b, \bm b_\sT^2)\, , \\
{\cal C}\left [w^{h^\perp}_{\scriptscriptstyle LT}\, h_{1\sL}^{g}\, h_{1 \sT}^{\perp\, g} \right] 
    & = \frac{M_h^5}{96\pi} \int_0^\infty \d |\bm b_\sT|\, \bm b_\sT^6\, J_1(|\bm b_\sT| |\bm q_\sT|)\, \tilde{h}_{1\sL}^{g\, (2)}(x_a, \bm b_\sT^2) \, \tilde{h}_{1\sT}^{\perp\, g\, (3)}(x_b, \bm b_\sT^2) \, , \\
{\cal C}\left[ w^f_{\scriptscriptstyle TT}\, f_{1 \sT}^{\perp\, g}\, f_{1 \sT}^{\perp\, g} \right] 
    & = - \frac{M_h^2}{4\pi} \int_{0}^{\infty} \d |\bm b_\sT| \, |\bm b_\sT|^3 \, J_0(|\bm b_\sT| |\bm q_\sT|) \, \tilde{f}_{1\sT}^{\perp\, g\, (1)}(x_a, \bm b_\sT^2) \, \tilde{f}_{1 \sT}^{\perp\, g\, (1)}(x_b, \bm b_\sT^2) \, , \\
{\cal C}\left[ w^g_{\scriptscriptstyle TT}\, g_{1 \sT}^{\perp\, g}\, g_{1 \sT}^{\perp\, g} \right] 
    & = - \frac{M_h^2}{4\pi} \int_{0}^{\infty} \d |\bm b_\sT| \, |\bm b_\sT|^3 \, J_0(|\bm b_\sT| |\bm q_\sT|) \, \tilde{g}_{1\sT}^{\perp\, g\, (1)}(x_a, \bm b_\sT^2) \, \tilde{g}_{1 \sT}^{\perp\, g\, (1)}(x_b, \bm b_\sT^2) \, , \\
{\cal C}\left[ w^h_{\scriptscriptstyle TT}\, h_{1}^{g}\, h_1^g \right] 
    & = -\frac{M_h^2}{8 \pi} \int_0^\infty \d |\bm b_\sT| \, |\bm b_\sT|^3\, J_0(|\bm b_\sT| |\bm q_\sT|) \, \tilde{h}_{1}^{g\, (1)}(x_a, \bm b_\sT^2)\, \tilde{h}_1^{g\, (1)}(x_b, \bm b_\sT^2)\, , \\
{\cal C}\left [w^{h^\perp}_{\scriptscriptstyle TT}\, h_{1\sT}^{\perp\, g}\, h_{1 \sT}^{\perp\, g} \right] 
    & = -\frac{M_h^6}{1152\pi} 
    \int_0^\infty \d |\bm b_\sT|\, |\bm b_\sT|^7\, J_0(|\bm b_\sT| |\bm q_\sT|)\, \tilde{h}_{1\sT}^{\perp\, g\, (3)}(x_a, \bm b_\sT^2) \, \tilde{h}_{1\sT}^{\perp\, g\, (3)}(x_b, \bm b_\sT^2) \, , \\
{\cal C}\left[ \overline{w}^f_{\scriptscriptstyle TT}\, f_{1 \sT}^{\perp\, g}\, f_{1 \sT}^{\perp\, g} \right] 
    & = \frac{M_h^2}{4\pi} \int_{0}^{\infty} \d |\bm b_\sT| \, |\bm b_\sT|^3 \, J_2(|\bm b_\sT| |\bm q_\sT|) \, \tilde{f}_{1\sT}^{\perp\, g\, (1)}(x_a, \bm b_\sT^2) \, \tilde{f}_{1 \sT}^{\perp\, g\, (1)}(x_b, \bm b_\sT^2) \, , \\
{\cal C}\left[ \overline{w}^g_{\scriptscriptstyle TT}\, g_{1 \sT}^{\perp\, g}\, g_{1 \sT}^{\perp\, g} \right] 
    & = \frac{M_h^2}{4\pi} \int_{0}^{\infty} \d |\bm b_\sT| \, |\bm b_\sT|^3 \,  J_2(|\bm b_\sT| |\bm q_\sT|) \, \tilde{g}_{1\sT}^{\perp\, g\, (1)}(x_a, \bm b_\sT^2) \, \tilde{g}_{1 \sT}^{\perp\, g\, (1)}(x_b, \bm b_\sT^2) \, , \\
{\cal C}\left[ \overline{w}^{h h^\perp}_{\scriptscriptstyle TT}\, h_{1}^{g}\, h_{1\sT}^{\perp\, g} \right] 
    & = -\frac{M_h^4}{96 \pi} \int_0^\infty \d |\bm b_\sT| \,|\bm b_\sT|^5\, J_2(|\bm b_\sT| |\bm q_\sT|) \, \tilde{h}_{1}^{g\, (1)}(x_a, \bm b_\sT^2)\, \tilde{h}_{1\sT}^{\perp\, g\, (3)}(x_b, \bm b_\sT^2)\, .
\label{eq: convolution h1T h1}
\end{align}

\providecommand{\href}[2]{#2}\begingroup\raggedright\endgroup

\end{document}